\documentclass[fleqn,usenatbib]{mnras}

\usepackage[T1]{fontenc}
\usepackage{ae,aecompl}
\usepackage[normalem]{ulem}

\usepackage{graphicx}	
\usepackage{amsmath}	
\usepackage{amssymb}	
\usepackage{subfig}
\usepackage{cuted}
\usepackage{stmaryrd}

\newcommand{\boldalpha}{\boldsymbol{\alpha}}
\newcommand{\hatn}{\hat{\mathbf{n}}}
\newcommand{\expval}[1]{\left\langle#1\right\rangle}

\usepackage{xcolor}

\usepackage{xspace}
\newcommand{\lowz} {\textsf{LOWZ}\xspace}
\newcommand{\cmass} {\textsf{CMASS}\xspace}
\newcommand{\qso} {\textsf{QSO}\xspace}
\newcommand{\LCDM}{$\Lambda$CDM\xspace}

\newcommand{\omegab}{\omega_\mathrm{b}}
\newcommand{\omegac}{\omega_\mathrm{c}}
\newcommand{\omegam}{\omega_\mathrm{m}}
\newcommand{\Omegab}{\Omega_\mathrm{b}}
\newcommand{\Omegac}{\Omega_\mathrm{c}}
\newcommand{\Omegam}{\Omega_\mathrm{m}}
\newcommand{\As}{A_\mathrm{s}}
\newcommand{\ns}{n_\mathrm{s}}
\newcommand{\zre}{z_\mathrm{re}}
\newcommand{\mnu}{{m_{\nu}}}
\newcommand{\Smnu}{{\Sigma m_{\nu}}}
\newcommand{\kcmb}{{\kappa_{\mathrm{CMB}}}}

\usepackage{physics}            
\let\oldvb\vb                   
\renewcommand{\vb}{\oldvb*}
\let\oldvu\vu
\renewcommand{\vu}{\oldvu*}

\usepackage[]{siunitx}          
\DeclareSIUnit \h {\ensuremath{\mathit{h}}}
\DeclareSIUnit \parsec {pc}
\DeclareSIUnit\lightyear{ly}
\DeclareSIUnit\year{yr}
\DeclareSIUnit\deg{deg}
\DeclareSIUnit\solarmass{\ensuremath{\textup{M}_\odot}}

\usepackage{etoolbox}
\makeatletter
\patchcmd\@combinedblfloats{\box\@outputbox}{\unvbox\@outputbox}{}{\errmessage{\noexpand patch failed}}
\makeatother

\usepackage{cleveref}

\title[Cosmological constraints from CMB and LSS]{Cosmological constraints from a joint analysis of cosmic microwave background and {spectroscopic tracers of the} large-scale structure}

\author[C. Doux et al.]{
Cyrille Doux,$^{1}$\thanks{E-mail: cdoux@apc.in2p3.fr}
Mariana Penna-Lima,$^{1,2,3}$
Sandro D. P. Vitenti,$^{2,4,5}$
\newauthor
Julien Tr\'eguer,$^{1}$
Eric Aubourg,$^{1}$
and Ken Ganga$^{1}$
\\
$^{1}$APC, AstroParticule et Cosmologie, Universit\'e Paris Diderot, CNRS/IN2P3, CEA/Irfu, Observatoire de Paris, Sorbonne Paris Cit\'e,\\ 10, rue Alice Domon et L\'eonie Duquet, 75205 Paris Cedex 13, France\\
$^{2}$Centro Brasileiro de Pesquisas F\'isicas, Rua Dr. Xavier Sigaud 150, CEP 22290-180, Rio de Janeiro, RJ, Brazil \label{inst2}\\ $^{3}$Laboratoire d'Annecy-le-Vieux de Physique des Particules (LAPP), Universit\'e Savoie Mont Blanc, CNRS/IN2P3, \\F-74941 Annecy, France \\
$^{4}$ Institut d'Astrophysique de Paris (${\cal G}\mathbb{R}\varepsilon\mathbb{C}{\cal O}$), UMR 7095 CNRS, \\ Sorbonne Universit\'es, UPMC Univ. Paris 06, Institut Lagrange de Paris, 98 bis boulevard Arago, 75014 Paris, France \\
$^{5}$Centre for Cosmology, Particle Physics and Phenomenology,
Institute of Mathematics and Physics, Louvain University, \\ 2 Chemin du Cyclotron, 1348 Louvain-la-Neuve, Belgium
}

\date{Accepted XXX. Received YYY; in original form ZZZ}

\pubyear{2017}

\begin{document}
\label{firstpage}
\pagerange{\pageref{firstpage}--\pageref{lastpage}}
\maketitle

\begin{abstract}
The standard model of cosmology, \LCDM, is the simplest model that matches the current observations, but it relies on two hypothetical components, to wit, dark matter and dark energy. Future galaxy surveys and cosmic microwave background (CMB) experiments will independently shed light on these components, but a joint analysis that includes cross-correlations will be necessary to extract as much information as possible from the observations. In this paper, we {{carry out} a multi-probe analysis based on pseudo-spectra} and test it on publicly available data sets. We use CMB temperature anisotropies and CMB lensing observations from Planck as well as the spectroscopic galaxy and quasar samples of SDSS-III/BOSS, taking advantage of the large areas covered by these surveys. We build a likelihood to simultaneously analyse the auto and cross spectra of CMB lensing and tracer overdensity maps before running Monte-Carlo Markov Chains (MCMC) to assess the constraining power of the combined analysis. We then add {the CMB temperature anisotropies likelihood} and obtain constraints on cosmological parameters ($H_0$, $\omegab$, $\omegac$, $\ln10^{10}\As$, $\ns$ and $\zre$) and galaxy biases. {We demonstrate that the joint analysis can additionally constrain the total mass of neutrinos $\Smnu$ as well as the dark energy equation of state $w$ at once (for a total of eight cosmological parameters), which is impossible with either of the data sets considered separately}. Finally, we discuss limitations of the analysis related to, \emph{e.g.}, the theoretical precision of the models, particularly in the non-linear regime.
\end{abstract}




\section{Introduction}
\label{sec:Introduction}

The large amount of cosmological data collected in the last few decades has been shedding light on the content of the Universe. Assuming General Relativity (GR) and the cosmological principle, the combination of different cosmological probes, such as type Ia supernov\ae, primary anisotropies of the cosmic microwave background (CMB), and large-scale structure (LSS) information, among others, indicates that the universe is almost flat, is dominated today by a dark energy (DE) component driving the current accelerated expansion phase of the Universe, and has some form of cold dark matter in addition to baryons and radiation \citep{2016A&A...594A..13P}. The flat \LCDM model is currently the simplest model compatible with the data of these combined probes.

We are reaching a precision era in cosmology and we may be able, in the near future, to distinguish between various cosmological models and achieve a better understanding of the fundamental nature of the DE and DM components.
Upcoming photometric and spectroscopic galaxy surveys such as the Large Synoptic Survey Telescope \citep[LSST,][]{2009arXiv0912.0201L}, Euclid \citep{2010arXiv1001.0061R}, the Wide Field Infrared Survey Telescope \citep[WFIRST,][]{2013arXiv1305.5422S} and the Dark Energy Spectroscopic Instrument \citep[DESI,][]{Levi:2013ul,2009arXiv0904.0468S}, aim at shedding light on those questions by probing the matter density field with ground-breaking precision. They will provide the data necessary for a deeper investigation of \LCDM and its competitors, hopefully allowing us to distinguish them.
Additionally, secondary anisotropies of the CMB due to gravitational lensing, the thermal (tSZ) and kinetic (kSZ) Sunyaev-Z'eldovich effects and the integrated Sachs-Wolfe (ISW) effect encode much information about dark matter and dark energy \citep{2000ApJ...540..605P}. Therefore, future CMB experiments, such as the Simons Observatory \citep{Suzuki:2016ca} and the Stage-IV CMB experiment \citep[CMB-S4,][]{2016arXiv161002743A}, will provide valuable complementary observations.

While various observations from multiple telescopes will provide exquisite and hopefully complementary data sets \citep{2015arXiv150107897J,2017ApJS..233...21R}, they will all observe the same sky, \emph{i.e.} the same underlying matter density field. Therefore, the observables they will measure are potentially statistically correlated. In this context, the cross-correlation between cosmological probes of different experiments yields new information, that is less prone to biases since different experiments are assumed to have uncorrelated noise and independent systematic effects. This correlation needs to be taken into account in the joint statistical analysis of multiple data sets in order to properly extract as much information as possible from it, without underestimating error bars on cosmological parameters \citep{2015APh....63...42R}. If this makes the analysis more demanding, the outcome is expected to provide stronger constraints on \emph{e.g.} dark energy, dark matter, the total mass of neutrinos \citep{2014PhRvD..89d3516P} or primordial non-gaussianities \citep{2012PhRvD..85d3518T}.

Initially, some of the best-explored cross-correlation information was that from CMB and galaxy surveys in order to measure the ISW signal \citep{1996PhRvL..76..575C,1998NewA....3..275B,2003ApJ...597L..89F,2006MNRAS.372L..23C,2016ApJ...826..121M}. But over the last decade, many different cross-correlation signals have been detected, combining various probes: the CMB anisotropies themselves, the CMB lensing potential, galaxy clustering, cosmic shear from the observations of galaxy weak lensing, etc. In particular, correlations of the gravitational lensing of the CMB with positions of galaxies \citep{2008PhRvD..78d3520H,2008PhRvD..78d3519H,2012ApJ...753L...9B,2012PhRvD..86h3006S,2013ApJ...776L..41G,2015ApJ...802...64B,2015ApJ...802L...1F,2015MNRAS.451..849A,2016MNRAS.461.4099B,2016MNRAS.456.3213G,2016MNRAS.460..434H,2016ApJ...825...24B} and lensing of galaxies \citep{2015PhRvD..91f2001H,2016MNRAS.459...21K} have been measured for various surveys. These measurements can provide unbiased estimates of galaxy biases, which encode the link between baryonic and dark matter, or the shear multiplicative bias \citep{2012ApJ...759...32V,2016PhRvD..93j3508L,2017PhRvD..95l3512S}. Finally, they have also been used to detect new signals, \emph{e.g.} the first detection of CMB lensing by cross-correlation with the NRAO VLA sky survey \citep{2007PhRvD..76d3510S}, the tSZ effect \citep{2013JCAP...11..064H,2014JCAP...02..030H}, the kSZ effect \citep{2012PhRvL.109d1101H,2016PhRvD..93h2002S},
and the position-dependent Lyman-$\alpha$ power spectrum \citep{2016PhRvD..94j3506D}.

{
Since cross-correlation signals are reaching high signal-to-noise ratio, joint analyses, \emph{i.e.} multi-probe analyses that exploit cross-correlations, are rapidly developing.
{In particular, joint analyses based on real-space correlation functions or power spectra have been used to test the consistency of cosmological constraints derived from different observations and to cross-calibrate nuisance parameters.}
\citet{2010A&A...523A...1J} forecasted that the joint analysis of galaxy weak lensing and galaxy density, including cosmic shear, galaxy clustering and the galaxy-galaxy lensing cross-correlation, could, at once, self-calibrate intrinsic alignments and constrain parameters of the cosmological model (see \citet{2016MNRAS.456..207K} as well).
{Since, several} analyses were
published that make use of the correlation between galaxy lensing and either galaxy density or CMB lensing {on available data}.
\citet{2016MNRAS.461.4099B} used SPT and DES-SV lensing data together with tracers of the large-scale structure and took advantage of the low systematic level of these {angular cross-}correlations {functions} to infer cosmological constraints, {demonstrating their consistency across data sets.}
\citet{2017MNRAS.464.4045K} used the galaxy clustering and galaxy-galaxy lensing signals of the DES-SV data {to obtain robust constraints on $\sigma_8$ and $\Omegam$.}
\citet{2017MNRAS.464.2120S} combined the lensing and clustering of SDSS galaxies with CMB lensing from Planck {to constrain clustering and shear biases and measure distance ratios, found to be consistent with Planck predictions}.
Very recent studies combined these three correlation functions in a single analysis: \citet{vanUitert:2017vu} with KiDS\footnote{The Kilo-Degree Survey, \citet{2013ExA....35...25D}.} and GAMA\footnote{Galaxies And Mass Assembly \citep{2009A&G....50e..12D}, a database of low-redshift surveys spanning the electromagnetic spectrum from radio waves to the ultraviolet domain.} galaxies, \citet{Joudaki:2017ui} with KiDS and BOSS galaxies (using the quadrupole of the power spectrum as well), \citet{2017arXiv170801530D} with the first year of DES data (forecasts were obtained in \citet{Park:2016fz} and the {robustness of the likelihood analysis pipeline was tested} in \citet{Krause:2017tj}), and \citet{2015ApJ...806....1M,2015ApJ...806....2M} used the overlapping area between CFHT and BOSS galaxies. Taking a step further, \citet{2016PhRvD..94h3517N,2017PhRvD..95h3523N} used, in a fully joint analysis, information from CMB temperature, CMB lensing, photometric surveys (both galaxy positions and lensing) and distance measurements from supernovae and direct $H_0$ measurements. Finally, \citet{2018arXiv180205257B} recently laid out the methodology for a joint analysis of five DES and SPT+Planck data two-point {auto- and cross-correlation} functions.
}

{
In this work, we aim at performing a joint analysis of Planck CMB data with the spectroscopic LSS tracers of the Baryon Oscillation Spectroscopic Survey \citep[SDSS-III/BOSS,][]{2013AJ....145...10D,2016MNRAS.455.1553R,2017A&A...597A..79P} {based on power spectra}, taking advantage of the large overlap between these surveys and the large redshift range of the BOSS samples, and developing an independent pipeline that is able to incorporate observables with different masks (thus maximizing signal-to-noise).
To this end, we build a Gaussian joint likelihood of {auto- and cross- pseudo power spectra of large-scale structure probes --here, CMB lensing and the contrast densities of tracers--}
and implement it as part of the public Numerical Cosmology\footnote{\url{http://numcosmo.github.io}} library \citep[\texttt{NumCosmo},][]{2014ascl.soft08013D}. In particular, we use the Planck 2015 lensing map \citep{2016A&A...594A..15P} and the three spectroscopic samples of BOSS --that is, {\lowz}, {\cmass} and the uniform quasar sample (\qso). We then run several Markov Chain Monte-Carlo (MCMC) analyses to extract constraints on cosmological parameters.
{We complete our analysis by adding the likelihood of the auto power spectrum of CMB temperature anisotropies from Planck \citep[thus neglecting the small CMB-LSS direct correlations sourced by the ISW effect,][]{2016A&A...594A..11P,2016A&A...594A..13P}, which allows us to perform a joint statistical analysis of CMB and LSS probes.} 
{We} demonstrate the performance of such an analysis to constrain the \LCDM model {at first, fitting the six base parameters ($H_0$, $\omegab$, $\omegac$, $\As$, $\ns$ and $\zre$), and then extensions of this base model, {fitting} the total mass of neutrinos, $\Smnu$, and the DE equation of state, $w=p/\rho${, separately and then jointly}. Finally, we perform the joint analysis using two different cut-offs at small scales in order to quantify their impact on cosmological constraints.
}

{
Note that our work follows an approach similar to that presented in \citet{2016PhRvD..94h3517N}{, which is also based on power spectra}. {However, our goals differ as we aim at constraining extensions of the \LCDM model, which we do using, in part, different data sets (we use spectroscopic data from BOSS) and we propose} a different formalism based on pseudo power spectra and a semi-analytical {covariance matrix evaluated at each set of cosmological parameters during the MCMC analysis, rather than a fixed covariance matrix (either simulation-based or computed analytically for a fiducial model)}. We also provide a new, public implementation of multi-probe analysis as part of the public \texttt{NumCosmo} library.  

}

The paper outline is as follows. In \cref{sec:Theory}, we develop the theoretical formalism, considering the Limber formula, to compute the pseudo angular power spectra of CMB lensing and galaxy overdensities. In \cref{sec:Data}, we describe the data used in this work and prepare the galaxy and quasar density maps. In \cref{sec:Methodology}, we develop the estimators and construct the likelihood used in this work, and then we perform the validation and null tests. In \cref{sec:analyses_and_results}, we detail the results of the statistical analyses and present constraints on cosmological parameters as well as their potential limitations related to theoretical uncertainties {(\emph{e.g.} related to the non-linear power spectrum)}, contamination of cross-spectra or data cuts. Finally, we draw conclusions in \cref{sec:conclusion}.

\section{Theoretical background}
\label{sec:Theory}

\subsection{{Angular power spectra}}
\label{sub:cross-correlation}

The matter distribution of the Universe is traced by cosmological probes such as galaxies, quasars (QSOs) and CMB lensing, among others. Their projected random fields on the observed direction $\hatn$ can be written as
\begin{equation}
	\label{eq:fieldA} A(\hatn) = \int_0^\infty dz \ W^A(z) \ \delta(\chi(z)\hatn, z),
\end{equation}
where $W^A(z)$ is the kernel function of an observable $A$ (kernels of the probes used in this work are plotted in \cref{fig:kernels}), $\delta = \delta\rho / \rho$ is the matter density contrast, $\rho$ is the matter density, and $\chi(z)$ is the comoving distance at redshift $z$. The fields $A(\hatn)$ and $B(\hatn)$, associated with galaxy or QSO density contrast and CMB lensing,
are assumed to be statistically homogeneous and isotropic.
Therefore the correlation function $\expval{A(\hatn)B^*(\hatn^\prime)}$ only depends on ${\hatn \cdot \hatn^\prime}$ and can be expanded as
\begin{equation}
    \expval{A(\hatn)B^*(\hatn^\prime)} = \sum_{\ell = 0}^{\infty} \frac{(2l + 1)}{4\pi} P_{\ell}(\hatn \cdot \hatn^\prime) C_{\ell}^{AB}
    \label{eq:cl_1}
\end{equation}
where $P_{\ell}$ are the Legendre polynomials, which defines the angular power spectrum $C_{\ell}^{AB}$.\footnote{Expanding the fields in spherical harmonics, $A{(\hatn)=\sum_{{\ell}m} A_{{\ell}m} Y_{{\ell}m}(\hatn)}$, this implies that ${\expval{A_{{\ell}m}B_{{\ell}^\prime m^\prime}} = C_{\ell}^{AB}\delta_{\ell\ell^\prime}\delta_{mm^\prime}}$.}

Using the inverse Fourier transform of the matter density field in eq.~\eqref{eq:fieldA} and substituting it into eq.~\eqref{eq:cl_1}, we obtain
\begin{align}
	\label{eq:cl_general} C_{\ell}^{AB} &= \int dz W^A(z) \int dz^\prime W^B (z^\prime) \\
	&\times \int dk \frac{2}{\pi} k^2 P(k, z, z^\prime) j_{\ell}(k\chi(z)) j_{\ell} (k\chi(z^\prime)), \nonumber
\end{align}
where $P(k, z, z^\prime)$ is the matter power spectrum and $j_{\ell}$ are the spherical Bessel functions. In this work, we adopt the Limber approximation \citep{1953ApJ...117..134L,2008PhRvD..78l3506L} and assume that spatial sections of the Universe are flat\footnote{It is worth noting that Limber approximation can be also applied for curved space universe, see \citet{2014JCAP...09..032L}, for instance.}, thus eq.~\eqref{eq:cl_general} becomes
\begin{equation}\label{eq:limber}
	C_{\ell}^{AB} = \int_0^{z_*} dz \frac{H(z)}{c \chi^2(z)} W^A(z) W^B (z) P\left(k = \frac{\ell +1/2}{\chi(z)}, z \right)
\end{equation}
where $c$ is the speed of light and $H(z)$ is the Hubble parameter. This approximation is valid when $P(k, z, z^\prime)$ varies slowly in comparison with the Bessel functions. In particular, the CMB lensing spectra are accurate for $\ell > 10$ \citep{2014JCAP...09..032L}. The selection functions of galaxies and quasars are wider than the largest scales probed\footnote{For a given sample, the largest scale probed is ${\chi_{\mathrm{max}} \sim \pi/k_{\mathrm{min}}}$ with ${k_{\mathrm{min}} = (\ell_{\mathrm{min}} + 1/2)/\chi(z_{\mathrm{eff}})}$, where $\chi(z_{\mathrm{eff}})$ is the comoving distance at the mean redshift of the sample and $\ell_{\mathrm{min}} = 20$, see \cref{sec:Data}. For {\lowz}, {\cmass} and {\qso}, these scales  are of order \SIlist{110; 220;630}{\per\h\mega\parsec}, while the selection functions have widths of order \SIlist{1080;860;970}{\per\h\mega\parsec}.} for the spectroscopic tracers used here. Therefore, we can safely make use of this approximation to compute theoretical power spectra, which are integrals of the matter power spectrum weighted by the kernel functions corresponding to each observable. We detail this in the following sections.

\begin{figure}
	\includegraphics[width=\columnwidth]{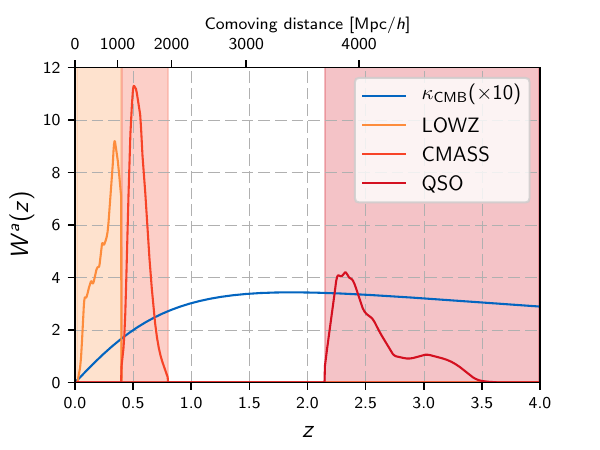}
	\caption{Kernel functions $W^A(z)$ of the observables used for cross-correlation as defined in equations~\eqref{eq:Wk} and \eqref{eq:Wg}. For the {\lowz}, {\cmass} and {\qso} samples, $W^g(z)$ reflects the redshift distribution (multiplied by the bias). The background colours correspond to the extent of the redshift distributions of the three samples. The CMB lensing kernel (multiplied by 10 on this plot for visibility) is very broad and peaks around $z\approx2$.}
	\label{fig:kernels}
\end{figure}

\subsection{Cosmic microwave background gravitational lensing}
\label{sub:Cosmic microwave background gravitational lensing}

The trajectories of the CMB photons are disturbed by the matter distribution such that, among other effects, the observed anisotropies of the temperature field in a direction $\hatn$ correspond to the unlensed field deflected by $\boldalpha$, i.e., $\tilde{T}(\hatn) = T(\hatn + \boldalpha)$ \citep{2006PhR...429....1L}. Assuming the small-angle Born approximation, $\boldalpha$ comprises the variations in the gravitational potential $\Psi$ along the line of sight from today's observer to the last scattering surface at redshift $z_*$, i.e.,
\begin{equation}
	\boldalpha = -2 \int_0^{z_*} dz \frac{c}{H(z)} \frac{\chi(z_*) - \chi(z)}{\chi(z)\chi(z_*)} \nabla_{\hatn} \Psi (\chi\hatn, z).
\end{equation}

The remapping of the CMB temperature anisotropies due to the weak lensing effect is described at lowest order by the convergence $\kcmb = -\frac{1}{2} \nabla_{\hatn} \boldalpha$. Given that on small angular scales $\nabla^2_{\hatn} \simeq \nabla^2$ \citep{2000ApJ...530..547J} and using the Poisson equation and eq.~\eqref{eq:fieldA}, we obtain the CMB convergence kernel
\begin{equation}\label{eq:Wk}
	W^\kcmb (z) = \frac{3}{2} \frac{\Omegam H_0^2}{c} \frac{(1+z)}{H(z)} \chi(z) \frac{\chi(z_*) - \chi(z)}{\chi(z_*)},
\end{equation}
where $\Omegam$ and $H_0$ are the present-day matter density and Hubble constant, respectively.

\subsection{Large-scale structure tracers}
\label{sub:Large-scale structure tracers}

Similarly to the CMB lensing, the galaxy or quasar overdensity in the direction $\hatn$ is a function of $\delta(\chi \hatn, z)$, namely
\begin{equation}
	g(\hatn) = \int_0^\infty dz \, W^g (z) \delta(\chi \hatn, z).
\end{equation}
The kernel $W^g (z)$ is given by \citep{2000ApJ...540..605P,2011PhRvD..84f3505B}
\begin{align}\label{eq:Wg}
	W^g (z) = b(z) \frac{dn}{dz} + \frac{3\Omegam}{2c} \frac{H^2_0}{H(z)} (1+ z) \, \chi(z) \, (5s-2) \, g(z)
\end{align}
where
\begin{equation}
	g(z) = \int_z^{z_{*}} dz^\prime \left( 1 - \frac{\chi(z)}{\chi(z^\prime)}\right) \frac{dn}{dz^\prime}.
\end{equation}
The function $b(z)$ is the linear bias relating the galaxy overdensity to the matter overdensity at large scales as $\delta_g (\chi\hatn, z) = b(z) \delta (\chi\hatn, z)$, and $dn/dz$ is the normalised redshift distribution of the tracers, which also contains the survey selection function. The second term in eq.~\eqref{eq:Wg} is due to the effects of gravitational lensing, with two opposing contributions -- the dilation of the apparent surveyed volume and the magnification bias effect for flux-limited samples\footnote{Lensed galaxies may appear brighter than they are and pass the luminosity threshold.}, where
\begin{equation}
	s= \left. \frac{d \log N(<m)}{dm} \right|_{m=m_{\max}}.
\end{equation}
$N(<m)$ denotes the cumulative count of objects with a magnitude smaller than $m$ and the derivative is estimated at the faint end of the catalog (\citet{2005ApJ...633..589S,2007PhRvD..76j3502H}). This term can be neglected when using \textsf{LOWZ} and \textsf{CMASS} samples, but it is relevant for quasars \citep{2013JCAP...12..029C}. Following~\citet{2005ApJ...633..589S}, we use $s_{\qso}=0.2$ throughout this analysis.

\subsection{Pseudo spectra} 
\label{sub:pseudo_spectra}

Many galaxy and CMB surveys cover only a fraction of the sky due to, for example, the limited field of view or galactic contamination, among others. In order to properly account for the partial sky coverage in the calculation of the angular power spectra, we define the mask function associated with the field $A(\hatn)$ as being $\mathcal{W}^A(\hatn)$ with value 1 if the direction $\hatn$ lies in the observed region, and 0 otherwise.

\begin{figure}
	\includegraphics[width=\columnwidth]{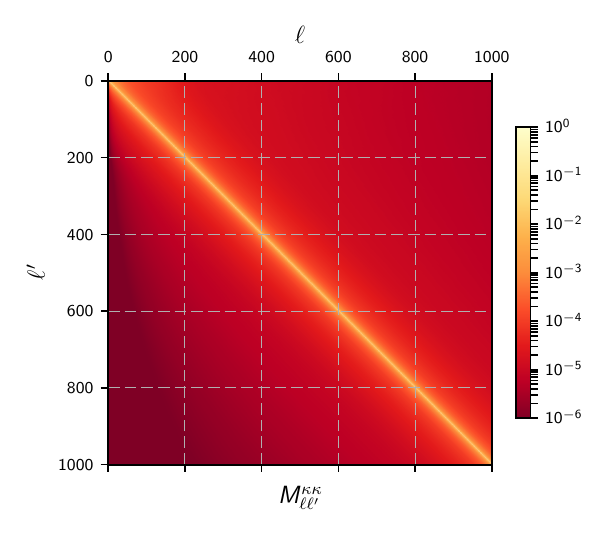}
	\caption{Mixing matrix of the CMB lensing auto power spectrum (see the mask in \cref{fig:kappaWF}) relating the full-sky power spectrum $C_{\ell}^{\kappa_{\mathrm{CMB}} \kappa_{\mathrm{CMB}}}$ to the pseudo spectrum $\tilde{C}_{\ell}^{\kappa_{\mathrm{CMB}} \kappa_{\mathrm{CMB}}}$ as in eq.~\eqref{eq:mean_ps}. The matrix elements are strongly dominated by the diagonal terms and the coupling between modes is given by off-diagonal elements. The other mixing matrices are qualitatively very similar.}
	\label{fig:mixkk}
\end{figure}

The cross-pseudo spectrum of observables $A$ and $B$ is thus defined as the cross spectrum of the cut-sky fields ${\tilde{A}(\hatn) = \mathcal{W}^A(\hatn) A(\hatn)}$ and ${\tilde{B}(\hatn) = \mathcal{W}^B(\hatn) B(\hatn)}$, and
its expected value can be related to the (true) full-sky cross spectrum in eq.~\eqref{eq:cl_general} by \citep{2005MNRAS.360.1262B}
\begin{equation}\label{eq:mean_ps}
	\langle \tilde{C}_{\ell}^{AB} \rangle = \sum_{\ell^\prime} M_{\ell \ell^\prime}^{AB} C_{\ell^\prime}^{AB},
\end{equation}
where $M_{\ell \ell^\prime}^{AB}$ is the mixing matrix which is given in terms of the Wigner-$3j$ symbols
\begin{equation}
	M_{\ell \ell^\prime}^{AB} = \frac{2\ell + 1}{4\pi} \sum_{\ell^{\prime\prime}} (2\ell^{\prime\prime}+1) \mathcal{W}_{\ell^{\prime\prime}}^{\, AB} \left(
	\begin{array}{ccc}
		\ell & \ell^\prime & \ell^{\prime\prime} \\
		0 & 0 & 0
	\end{array}
	\right)^2.
\end{equation}
The cross spectra of the masks are
\begin{equation}
\mathcal{W}_{\ell^{\prime\prime}}^{\, AB} = \frac{1}{2\ell^{\prime\prime} +1}\sum_m w^A_{\ell^{\prime\prime} m} (w^B_{\ell^{\prime\prime} m})^*,
\end{equation}
where
\begin{equation}
w^A_{\ell m} = \int d\hatn \mathcal{W}^A(\hatn) Y_{\ell m}^*.
\end{equation}

The mixing matrix introduces a scaling factor equal to $\left(f_{\rm sky}^A f_{\rm sky}^B\right)^{1/2}$, i.e., the geometric mean of the observed sky fractions for the observables $A$ and $B$, respectively, since the form of the masked function is constant \citep{2002ApJ...567....2H}. It also couples the multipoles $\ell$ and $\ell'$, that would be otherwise uncorrelated, especially at large scales. It is computed analytically for each pair of observables (see \cref{fig:mixkk}).

\section{Data}
\label{sec:Data}

\begin{figure*}
    \centering
    \subfloat[kappa][CMB lensing convergence.]{\includegraphics[width=0.5\textwidth]{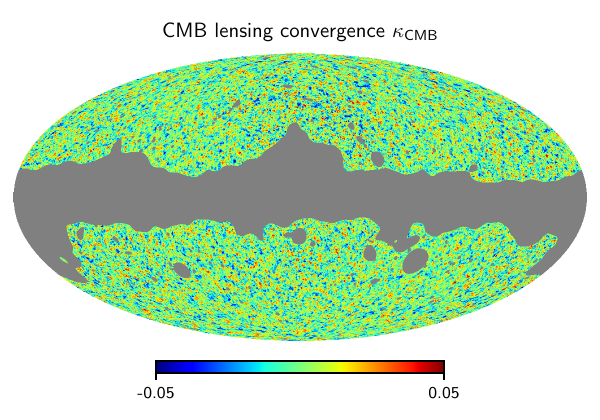}\label{fig:kappaWF}}
    \subfloat[lowz][LOWZ overdensity.]{\includegraphics[width=0.5\textwidth]{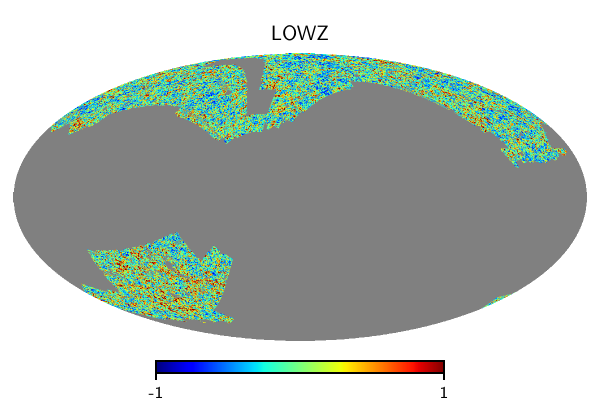}\label{fig:LOWZ_map}}

    \subfloat[cmass][CMASS overdensity.]{\includegraphics[width=0.5\textwidth]{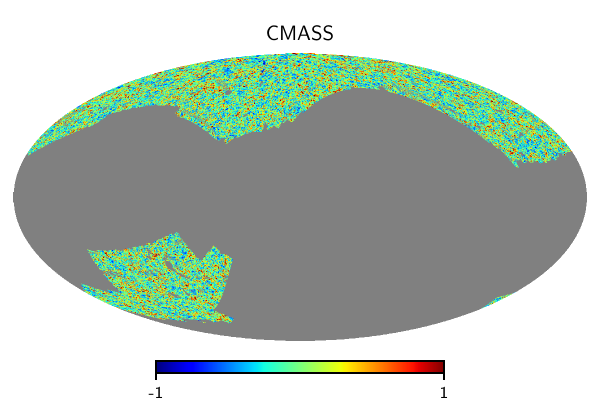}\label{fig:CMASS_map}}
    \subfloat[qso][QSO overdensity.]{\includegraphics[width=0.5\textwidth]{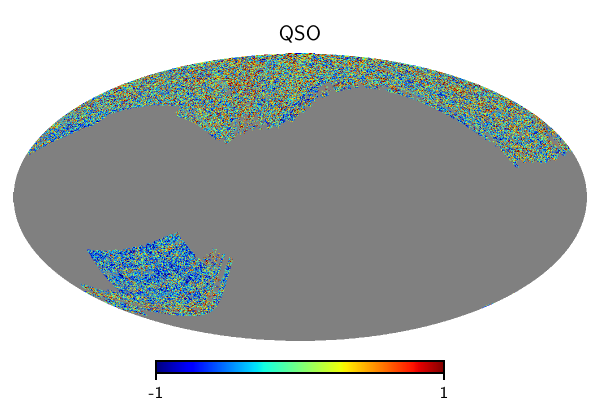}\label{fig:QSO_map}}
    \caption{Planck CMB lensing convergence and BOSS galaxy and quasar overdensity maps in galactic coordinates. The auto and cross spectra of these maps are used in this work. Grey areas correspond to the masked areas near the galactic plane. The lensing map has been Wiener-filtered and the overdensity maps have been smoothed on one degree scale for visualisation purposes only.}
    \label{fig:maps}
\end{figure*}

\subsection{Planck data}
\label{sub:Planck_data}

{
Planck\footnote{\url{http://sci.esa.int/planck/}} \citep{2016A&A...594A...1P}, the fourth satellite to survey the CMB over the full sky, was launched on May 14\textsuperscript{th}, 2009. Its scientific payload comprised two instruments: the Low Frequency Instrument \citep{2010A&A...520A...3M}, which observed for four years in bands at 30, 44 and 70 GHz, and the High Frequency Instrument \citep{2010A&A...520A...9L}, which observed for almost two-and-a-half years in bands at 100, 143, 217, 353, 545 and 857 GHz.
}

{
In this work, we use Planck data for both the primordial CMB temperature anisotropies \citep{2016A&A...594A...9P} and the CMB lensing \citep{2016A&A...594A..15P}. For temperature anisotropies, we use the two Planck likelihood codes: \texttt{Plik} for the high multipoles, $\ell \geq 30$, and \texttt{Commander} for low multipoles, $\ell < 30$ \citep[see][for details on components separation]{2016A&A...594A..11P}. 
}

{
We also use the CMB lensing convergence map from the Planck 2015 data release \citep{2016A&A...594A..15P}. The Planck Collaboration provides the convergence map\footnote{The convergence map and mask files are publicly available at \url{http://pla.esac.esa.int/pla/}.} in the \texttt{Healpix}\footnote{\url{http://healpix.jpl.nasa.gov/}} \citep{2005ApJ...622..759G} format, with resolution parameter $N_{\rm side}=2048$, and the corresponding binary mask, with a sky fraction $f_{\rm sky} = 0.67$. Lensing potential maps were reconstructed from foreground-cleaned temperature and polarisation maps, obtained from the \texttt{SMICA} code.
These were used to form five quadratic estimators $\hat{\phi}^{TT}$, $\hat{\phi}^{TE}$, $\hat{\phi}^{EE}$, $\hat{\phi}^{EB}$ and $\hat{\phi}^{TB}$, combined into a minimum-variance estimator \citep[see ][for specificities about the Planck reconstruction]{2016A&A...594A..15P}. The Wiener-filtered convergence map is shown in \cref{fig:kappaWF} with its mask.
}

{
The full-sky lensing power spectrum was evaluated by the Planck collaboration following the methods detailed in Appendix~A of \citet{2016A&A...594A..15P}, which we briefly describe now: first, the pseudo power spectrum of the masked, reconstructed lensing map (itself using optimized temperature map masks) is upweighted by $1/f_{\rm sky}$. Then, different contributions must be subtracted: the Gaussian noise from the disconnected part of the four-point function (the N0 term), the non-Gaussian noise from the connected contribution (the N1 term) and the subdominant contribution of shot-noise from unresolved point sources (the PS term). Finally, a term (the MC term) derived from the mismatch between input and output power spectra derived from simulations corrects for errors in the normalization, mask-related mode mixing and computation of the N1 term. Finally, the likelihood of the binned, full-sky power spectrum includes small, linear corrections to take into account the dependence of the reconstructed lensing map and the N1 term on the fiducial cosmological model. 
}

{
Given our goal to jointly analyze CMB lensing with tracers of the LSS, we cannot simply use the lensing likelihood code provided by the Planck Legacy Archive. However, an estimate of the unbinned (pseudo) lensing auto-power spectrum has not been released yet. Therefore, considering that redoing the Planck analysis goes well beyond the scope of this paper, we measured the lensing pseudo-power spectrum from the released map and estimated the dominant N0 and N1 noise terms from the simulated reconstructed maps (see \cref{subs:pseudo_spectra_estimator}) and neglected the small MC term, the largely subdominant PS term and fiducial model correction terms. Indeed, the MC term partly corrects for differences between the full-sky and pseudo-power spectra while we only rely on the pseudo spectrum, and the corrections applied at the likelihood level are very small in the multipole ranges used for cosmology. Therefore, while a simplification, our method should provide reasonable results.
{Indeed, in \cref{sec:lensing_auto_power_spectrum}, we compare cosmological constraints obtained from CMB temperature and lensing using our pipeline and that provided by the Planck collaboration and demonstrate that we retrieve unchanged cosmological constraints, except for a $0.5\sigma$ shift in the $\As-\zre$ degeneracy, still below the statistical error. While the reader should bear in mind this small difference, we consider this comparison to justify our simplifications given the purposes of this work.}
}

\subsection{SDSS-III/BOSS data}
\label{sub:SDSS-III/BOSS data}

{The spectroscopic samples from the Baryon Oscillations Spectroscopic Survey \citep[BOSS,][]{2013AJ....145...10D} consists of two galaxy catalogs named \textsf{LOWZ} and \textsf{CMASS} and one quasar catalog, a subset of which has a uniform selection function.} They are extensively described in \citet{2016MNRAS.455.1553R} and we only summarize relevant information in this section.

\subsubsection{Luminous Red Galaxies: \textsf{LOWZ} \& \textsf{CMASS}}
\label{subs:LOWZ/CMASS}

\textsf{LOWZ} contains Luminous Red Galaxies (LRG) at low redshift ($z \lesssim 0.4$), and it aims at a constant number density of about $\bar{n} \sim \SI{3e-4}{\h\cubed\per\mega\parsec\cubed}$ over the redshift range $[0.1, 0.4]$. This is done using a redshift dependent magnitude cut. In this work, we use the twelfth data release (DR12) and select galaxies in the range {0-0.4}, which contains \num{383876} galaxies. The \textsf{CMASS} sample contains galaxies at higher redshifts $0.4 \lesssim z \lesssim 0.8$ with a constant stellar mass in this redshift range. The twelfth data release contains \num{849637} galaxies in the redshift range {0.4-0.8} used in this work. The normalised redshift distributions of the two samples are shown on \cref{fig:kernels} (multiplied by their respective biases).

BOSS's spectroscopic fibres are plugged into tiles of diameter \SI{3}{\degree} to observe predetermined targets. The combined footprints of all tiles can be decomposed into non-overlapping sky sectors. Because of the finite size of fibres, galaxies closer than 62' may not be observed even after multiple observations of the same field. The pipeline may also fail in determining the redshift of some galaxies (especially the faintest ones). Therefore, for each sector~$i$, the completeness is defined as the ratio of observed galaxies with a measured redshift to the number of targets lying in that same sector
\begin{equation}
	C_i = \frac{N_{\mathrm{obs, i}}} {N_{\mathrm{targ, i}}}.
\end{equation}
The completeness maps are defined in the \texttt{Mangle} software\footnote{See \url{http://space.mit.edu/~molly/mangle/}.} format and are converted to \texttt{Healpix} maps with resolution parameter $N_{\rm side}=2048$. The mask functions of the galaxy samples are obtained by assigning 1 to pixels where the completeness is above 75\% and then removing small areas that were vetoed for bad photometry, bright objects and stars and instrumental constraints, such as fibre centerposts and fibre collisions.

In order to correct for completeness, each galaxy is thus assigned a weight
\begin{equation}
	w_{\mathrm{tot}} = w_{\mathrm{star}} w_{\mathrm{seeing}} (w_{\mathrm{cp}} + w_{\mathrm{noz}} - 1),
\end{equation}
where $w_{\mathrm{star}}$ and $w_{\mathrm{seeing}}$ correct for non-cosmological fluctuation in the target selection due to stellar density (only for the \textsf{CMASS} sample) and atmospheric seeing, $w_{\mathrm{cp}}$ corrects for fibre collisions and $w_{\mathrm{noz}}$ corrects for redshift failures.

\subsubsection{Quasars}
\label{subs:QSO}

The selection function (over the sky) of the full quasar sample of BOSS is not uniform due to the observing strategy, hence we shall use the so-called CORE sample which contains QSOs with redshift $z \geq 2.15$ that were uniformly selected by the \texttt{XDQSO} algorithm \citep{2011ApJ...729..141B}. There are 94971 quasars in the CORE sample of DR12 within this redshift range. The completeness is computed using the \texttt{BOSSQSSOMASK} software\footnote{See \url{http://faraday.uwyo.edu/~admyers/bossqsomask/}.} from~\citet{2015MNRAS.453.2779E} and is then combined with the veto mask to build the mask of the quasar density map.

\subsubsection{Building the maps} 
\label{ssub:building_the_maps}

For the \textsf{LOWZ} and \textsf{CMASS} samples, we build \texttt{Healpix} maps with resolution parameter $N_{\rm side}=2048$, where for each pixel~$p$,
\begin{equation}
	\delta_p = \frac{N_{w(p)}}{\overline{N}} - 1.
	\label{eq:density}
\end{equation}
$N_{w(p)} = \sum_{i \in p} w_i$ is the number of galaxies in pixel $p$ counted with their weights and $\overline{N} = \frac{1}{N_{\rm pix}} \sum_{p=1}^{N_{\rm pix}} N_{w(p)}$ is the mean pixel count (where the sum runs only on pixels in the observed area, \emph{i.e.} where the mask function is equal to 1).

For the quasars, there is no weight provided in the BOSS DR12 catalog and the density map is computed as
{\begin{equation}
	\delta_p = \frac{\flatfrac{N_p}{C_{s(p)}}}{\overline{N}} - 1,
\end{equation}
where $N_p$ denotes the number of QSOs lying in pixel $p$, $C_{s(p)}$ is the completeness of the sector $s(p)$ where the pixel $p$ lies and $\overline{N}$ denotes the mean pixel count (up-weighted by completeness) in the observed area.}

The angular densities of the samples are $\bar{n}_{\lowz} = \SI{150e3}{\per\steradian}$, $\bar{n}_{\cmass} = \SI{300e3}{\per\steradian}$ and $\bar{n}_{\qso} = \SI{36e3}{\per\steradian}$. The maps of the estimated overdensity for the three samples are shown in \cref{fig:LOWZ_map,fig:CMASS_map,fig:QSO_map}.


\subsection{CMB lensing--large-scale structure correlations data} 
\label{sub:cross_correlation_data}

\begin{figure*}
\centering
	\includegraphics[width=17cm]{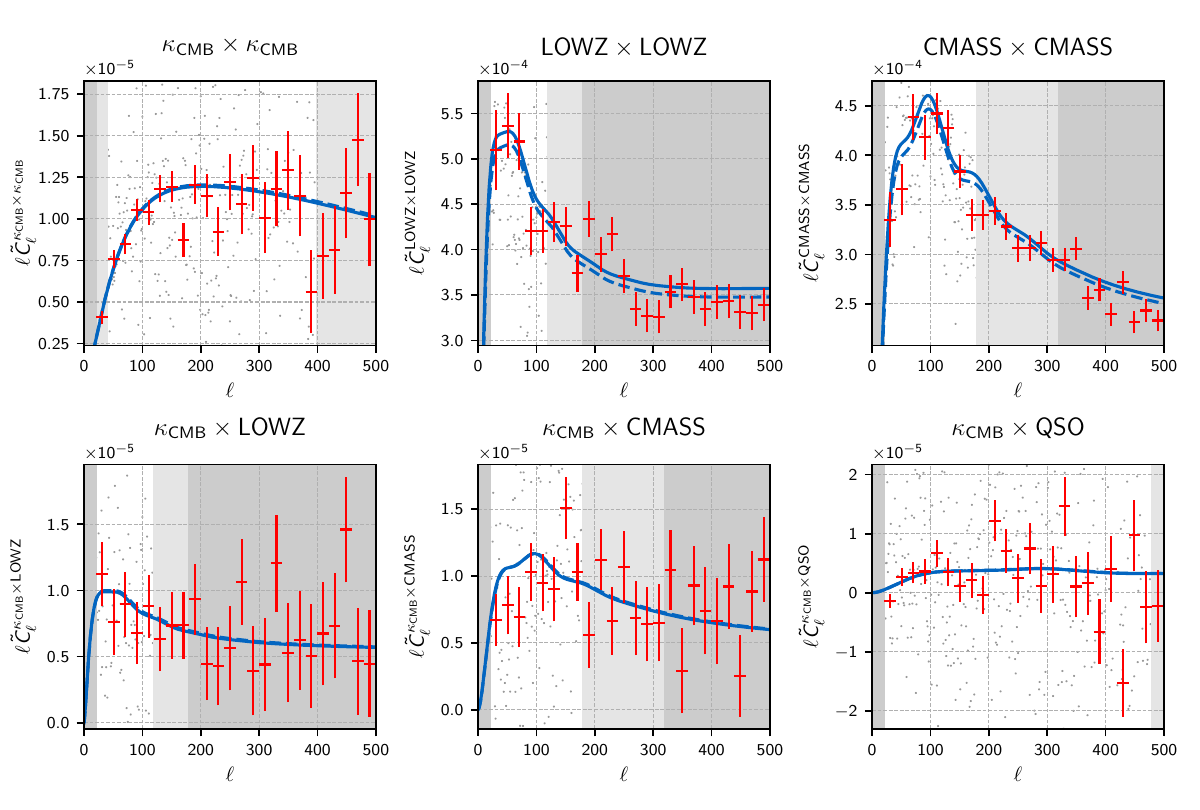}
	\caption{Auto- and cross-pseudo spectra used in this paper for the joint cosmological analysis of CMB lensing and spectroscopic tracers. Observed spectra are represented by the light grey points in the multipole range ${20-500}$, and binned as red error bars (only for visualisation). Theoretical curves {are shown in blue for the best-fit from the joint analysis on \LCDM with the optimistic cut (solid lines, see values in \cref{tab:constraints}) and for best-fit biases using a fixed Planck {2015} cosmology~\citep{2016A&A...594A..13P} (dashed lines) on the full multipole range}. Multipole ranges discarded in the cosmological analysis are shaded in grey (light grey shows the conservative cut, dark grey the optimistic one). Pseudo spectra are multiplied by $\ell$ to help visualise features of the theoretical power spectra, especially the wiggling related to baryon acoustic oscillations (covariance matrices are modified accordingly).}
	\label{fig:clobs}
\end{figure*}

In this work, we will use, in addition to CMB temperature data, auto and cross spectra of CMB lensing from Planck and spectroscopic tracers from BOSS. More precisely, we will use the auto-pseudo spectra of the CMB lensing map $\kappa_\mathrm{CMB}$ and of the density contrast maps of the {\lowz} and {\cmass} samples. We also use the pseudo-cross-spectra of the CMB lensing map with the three LSS tracers. The collection of these six spectra (shown in \cref{fig:clobs}) will henceforth be referred to as ``\emph{CMB lensing-LSS correlations}'', and denoted {``CMB lensing $\otimes$ BOSS tracers''} in the figures.

We do not use the auto spectrum of the {\qso} map because it is completely shot-noise-dominated in multipole space. We do not use the galaxy cross spectra because their redshift ranges do not overlap and the cross spectra should therefore be zero in the Limber approximation (which we check in the next section).

{We use different multipole ranges for the different spectra and describe here the cuts that were applied.
{At very large scales, the Limber approximation breaks \citep{2008PhRvD..78l3506L}, RSD becomes non-negligible \citep{Padmanabhan:2007ci} and observational systematics become more difficult to handle (see \cref{sub:null_tests}). For these reasons, we chose a minimum multipole of ${\ell_{\rm min}=20}$ common to all spectra involving LSS tracers.}
At small scales, uncertainties in the non-linear power spectrum severely constrain the use of angular power spectra. We therefore considered two cut-offs in Fourier space, using a fixed smallest scale $k_{\rm max}$ that translates into a maximum multipole $\ell_{\rm max}\approx\flatfrac{k_{\rm max}}{\chi\left(z_{\rm eff} \right)}$ where $\chi\left(z_{\rm eff} \right)$ is the comoving distance to the mean redshift of the samples. We will use a conservative cut-off at $k_{\rm max}=\SI{0.1}{\per\mega\parsec}$ and a more optimistic one at $k_{\rm max}=\SI{0.15}{\per\mega\parsec}$. For \lowz, \cmass and \qso, we find respectively $\ell_{\rm max}=\numlist{120;200;480}$ and $\ell_{\rm max}=\numlist{180;320;720}$. At these scales, the updated \texttt{halofit} model reaches close to percent-level precision, even for models including massive neutrinos \citep{2003MNRAS.341.1311S,2012MNRAS.420.2551B,2012ApJ...761..152T}. Moreover, the scale-independent bias approximation has been found to work reasonably well for scales down to $\SI{20}{\per\h\mega\parsec}$, with variations of less than 5\% \citep{2009MNRAS.392..682C,2016MNRAS.460.1173R}. We shall therefore, for both cuts, be in the regime where these assumptions are safe.
}

{
The CMB lensing power spectrum was measured in the range $\ell=8-2048$ \citep{2014A&A...571A..17P}, but due to potential errors in modelling large-scale survey anisotropies and Gaussian noise at small scales (the N0 disconnected component), we only consider multipoles $\ell=40-400$ and $\ell=20-500$ for the conservative and optimistic cuts. The high-$\ell$ cut is justified by the facts that Planck, as of the 2015 results, has the best sensitivity among CMB experiments at $\ell<500$ and that it includes additional signal-to-noise while discarding the range $500<\ell<700$ where the estimated power spectrum is significantly lower that expected for the \LCDM model. Uncertainties in the N1 term becomes more important at $\ell\gtrsim600$ and we shall therefore also be in a safe regime for both cuts considered here.
}

The six observed pseudo spectra are shown in \cref{fig:clobs} together with theoretical curves for our best-fit biases ($b_{\lowz}=1.831\pm0.048$, $b_{\cmass}=2.077\pm0.029$ and $b_{\qso}=2.21\pm0.44$) and with fixed cosmology.\footnote{Parameters' values are fixed at the best-fit cosmology for Planck ``TT,TE,EE+lowP+lensing+ext'' \citep{2016A&A...594A..13P} for the flat \LCDM model with a total mass of neutrinos {${\Smnu} = \SI{0.06}{\eV}$}.} We report detections for the CMB lensing-galaxy density cross-correlations of 4.7~$\sigma$, 13.9~$\sigma$ and 10.6~$\sigma$ for {\lowz}, {\cmass} and {\qso}, respectively.

\section{Methodology}
\label{sec:Methodology}

{This section describes the method to analyse the data. We choose to use the pseudo-power spectrum formalism, which takes into account partial sky coverage by  appropriately scaling the theoretical full-sky power spectra and comparing them to the observed pseudo-spectra. This forward-modelling method has the advantage that it provides simple ways to obtain unbiased prediction, without the need to reverse the effects of the mask on the observation, an operation that can be difficult and unstable given the complexity of the masks. Moreover, it naturally deals with observables using different masks, thus maximizing the signal-to-noise ratio of the Fourier coefficients of cross-spectra. The drawback is that the covariance matrix is impossible to compute analytically, even in the Gaussian field approximation: instead, we must either estimate it with Monte-Carlo simulations, or, as we do here, use a semi-analytical approximation. We perform several tests to validate this method and the statistical pipeline used in the next section. Finally, we also perform null tests searching for possible contamination of the power spectra by effects due to the masks or the SDSS photometry.
}

\subsection{Likelihood}
\label{sub:Likelihood}

\subsubsection{Pseudo spectra estimator}
\label{subs:pseudo_spectra_estimator}

Pseudo spherical harmonic coefficients $\tilde{A}_{\ell m}$ and $\tilde{B}_{\ell m}$ (for $A,B\in\{\lowz,\cmass,\qso,\kcmb\}$) of the four maps are estimated with the \texttt{map2alm} function of \texttt{Healpix}, {corrected for the \texttt{Healpix} pixel window function} and summed to give an estimator of the pseudo spectra
\begin{equation}
	\hat{\tilde{C}}_{\ell}^{AB} = \frac{1}{2\ell+1}\sum_{m=-\ell}^{m=+\ell} \tilde{A}_{\ell m} \tilde{B}_{\ell m}^*.
\end{equation}
These pseudo spectra have a noise contribution and an expectation value
\begin{equation}
	\langle \hat{\tilde{C}}_{\ell}^{AB} \rangle = \sum_{\ell'} M_{\ell \ell'}^{AB} C_{\ell'}^{AB} + \delta_{AB} \tilde{N}_{\ell}^{A}
\end{equation}
where $\tilde{N}_{\ell}^{A}$ is the noise pseudo-spectrum of the measured field $A(\hatn)$, which needs to be subtracted. It is assumed here that different observables have uncorrelated noise, \emph{i.e.} that noise cross spectra are null (for both full-sky and pseudo).

In principle, noise pseudo spectra can be computed using the mixing matrix and eq.~\eqref{eq:mean_ps}.
\begin{equation}
	\tilde{N}_{\ell}^{A} = \sum_{\ell'} M_{\ell \ell'}^{AA} N_{\ell'}^{A}.
\end{equation}
However, the sum over $\ell'$ runs from 0 to infinity, so, in practice, it has to be cut at some maximum multipole $\ell_{\max}$. But convergence is not guaranteed, since noise spectra are increasing functions of $\ell$ (they are quasi-constant for galaxies and grow like $\sim \ell^2$ for CMB lensing). Therefore, we used instead simulated noise maps for {CMB lensing and a shuffling technique for spectroscopic tracers positions (similar to that used in \citet{2016PhRvD..94h3517N})}.

For CMB lensing, we used the 100 simulated lensing reconstruction maps provided by the Planck Legacy Archive.\footnote{See \url{http://pla.esac.esa.int/pla/}.} Given a known, full-sky, input convergence map (to be masked) and a masked, reconstructed convergence map, one can compute the difference of the pseudo spectra in order to obtain an estimate of the noise pseudo power spectrum $\tilde{N}_{\ell}^{\kappa}$, which is then averaged over realisations of the simulation.


{For the clustering of spectroscopic tracers, the full-sky shot noise spectrum is constant, equal to $N_{\ell} = 1 / \bar{n}$ where $\bar{n}$ is the angular density of objects (weighted and expressed in steradian\textsuperscript{-1}). However, weights associated with each object in the galaxy spectroscopic samples to compensate incompleteness slightly increase the noise level (by up to 8\%)\footnote{This can be intuited by noting that observing two galaxies with weight 1 as \emph{more information} that observing one galaxy with weight two. In \citet{2012ApJ...761...14H}, this was dealt with by adding an additive free term.}. Therefore, we randomly reposition objects within the masks, keeping their weights. This operation breaks the spatial, cosmological correlation and therefore the cosmological contribution to the spectrum, leaving only Poisson noise with appropriate weighting. Density maps are then built according to the procedure described in \cref{ssub:building_the_maps} and their pseudo-spectra $\tilde{N}_{\ell}^{g}$ are evaluated with \texttt{Healpix}. This process is repeated one thousand times and the noise spectrum is averaged over realizations. The high resolution of the density maps allows us to measure the angular spectra at very large multipole values (up to $\ell=3N_{\rm side}-1=6143$) where the cosmological signal becomes negligible with respect to shot-noise. We find excellent agreement of our noise estimator and the one measured on the real density map. Moreover, we validate this process by fitting the measured spectra with a free additive constant as in \citet{2012ApJ...761...14H}, found to be consistent with zero, within $1\sigma$ error bars.}

Our estimator thus reads
\begin{equation}
	\hat{\tilde{C}}_{\ell}^{AB} = \frac{1}{2\ell+1}\sum_{m=-\ell}^{m=+\ell} \tilde{A}_{\ell m} \tilde{B}_{\ell m}^* - \delta_{AB}\frac{1}{N_{\rm sim}} \sum_{i=1}^{N_{\rm sim}} \hat{\tilde{N}}_{\ell}^{A,i},
\end{equation}
where $\hat{\tilde{N}}_{\ell}^{A,i}$ is the estimated pseudo-noise spectrum of simulation number $i$. The pseudo spectra used in this work are shown in \cref{fig:clobs}.

\subsubsection{Covariance matrix and likelihood} 
\label{ssub:covariance_matrix_and_likelihood}

\begin{figure}
	\includegraphics[width=\columnwidth]{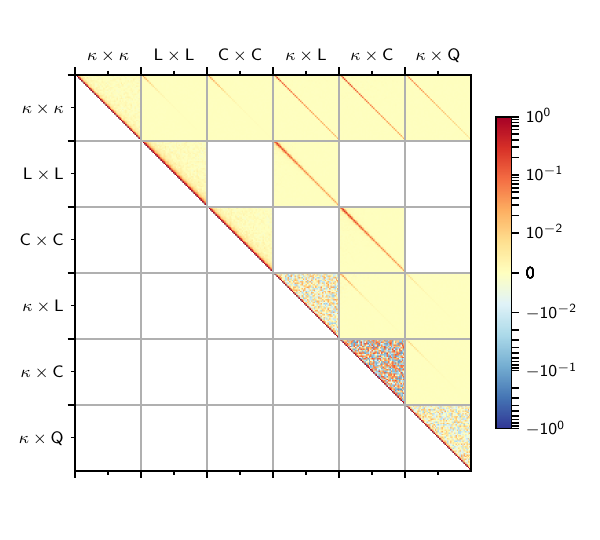}
	\caption{Full covariance matrix (normalised to unit diagonal) of the CMB lensing-LSS correlations computed from eq.~\eqref{eq:covariance} in symmetric logarithmic scale. It is divided in blocks corresponding to the six angular spectra: $\kappa$, L, C and Q correspond to respectively $\kappa_{\mathrm{CMB}}$, {\lowz}, {\cmass} and {\qso}. Note the (small) numerical noise in the variance blocks of the cross power spectra from the $\mathbf{X}/\mathbf{Y}$ matrices. Only the upper part is displayed. The white blocks in the upper parts correspond to non-correlated spectra.}
	\label{fig:cov}
\end{figure}

The covariance matrix of the pseudo spectra {used in this work assumes that the density field is Gaussian for the scales exploited in the analysis. It} is computed using an extension of Efstathiou's symmetrisation approximation \citep{2004MNRAS.349..603E} following \citet{2005MNRAS.360.1262B} and is given by
\begin{align}
	\mathrm{Cov}\left( \tilde{C}_{\ell}^{AB}, \tilde{C}_{\ell'}^{CD} \right) = & \sqrt{ D_{\ell}^{AD} D_{\ell'}^{AD} D_{\ell}^{BC} D_{\ell'}^{BC}} \mathbfss{X}_{\ell \ell'}^{ABCD}\nonumber \\
	&+ \sqrt{ D_{\ell}^{AC} D_{\ell'}^{AC} D_{\ell}^{BD} D_{\ell'}^{BD}} \mathbfss{Y}_{\ell \ell'}^{ABCD}
	\label{eq:covariance}
\end{align}
with
\begin{equation}
    D_{\ell}^{AB} = \left\{
    \begin{array}{ll}
        C_{\ell}^{AB} & \text{if } A \neq B \\
        C_{\ell}^{AA} + N_{\ell}^{A} & \text{if } A=B \\
    \end{array}\right. ,
\end{equation}
where $C_{\ell}^{AB}$ and $N_{\ell}^{A}$ are the full-sky theoretical and noise spectra. $\mathbfss{X}_{\ell \ell'}^{abcd}$ and $\mathbfss{Y}_{\ell \ell'}^{abcd}$ are two matrices depending only on the masks of observables ${A,B,C,D}$, determined to arbitrary precision by a Monte-Carlo (MC) simulation (see \cref{sec:XY_matrices_in_the_covariance} for more details). The covariance matrix (estimated for a fiducial cosmology) is shown in \cref{fig:cov}.

A Gaussian likelihood is used for the stacked pseudo spectra vector
\begin{equation}\label{eq:likelihood}
	\mathcal{L}\left(\tilde{\bmath{C}}_{\ell}^{\mathrm{obs}} \vert b_g, \Theta_{\mathrm{cosmo}} \right) =
	{\frac{1}{({2\pi)^{n/2}}{\lvert \mathbfss{Cov} \rvert}^{1/2}}}
	e^{-\chi^2 / 2},
\end{equation}
where
\begin{equation}
	\chi^2 = {\left( \tilde{\bmath{C}}^{\mathrm{obs}} - \tilde{\bmath{C}}^{\mathrm{th}} \right)^{\mathrm{T}}}
	\left[\mathbfss{Cov}\right]^{-1}
	{\left( \tilde{\bmath{C}}^{\mathrm{obs}} - \tilde{\bmath{C}}^{\mathrm{th}} \right)},
\end{equation}
$\tilde{\bmath{C}}^{\mathrm{obs}}$ is the stacked vector of observed pseudo spectra (see \cref{fig:clobs}) and $\tilde{\bmath{C}}^{\mathrm{th}}$ is the stacked vector of theoretical pseudo spectra computed from the Limber approximation (see eq.~\eqref{eq:limber}) and multiplied by the mixing matrices. The covariance matrix $\mathbfss{Cov}$ is that of the stacked vector as defined in eq.~\eqref{eq:covariance} and is shown in \cref{fig:cov}.


\subsection{Validation}
\label{sec:validation}

In this section, we perform validation tests for the pseudo spectrum estimator, the covariance matrix and the statistical pipeline.

\begin{figure}
	\includegraphics[width=\columnwidth]{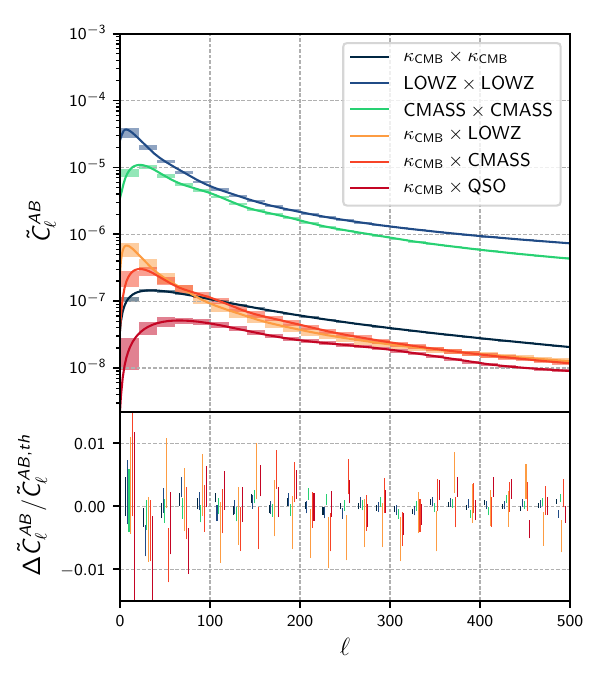}
	\caption{Validation of the pseudo spectrum estimator: the upper plot shows in solid line the theoretical pseudo spectra, computed using full-sky spectra and mixing matrices as in eq.~\eqref{eq:mean_ps}. Boxes show the mean of the simulated pseudo spectra and its spread for 1000 realisations, binned for visualisation. The lower plots shows the relative error. All spectra are consistent with the theoretical expectations in the multipole range used for this work.}
	\label{fig:pcl_validation}
\end{figure}

\begin{figure*}
\centering
	\includegraphics[width=17cm]{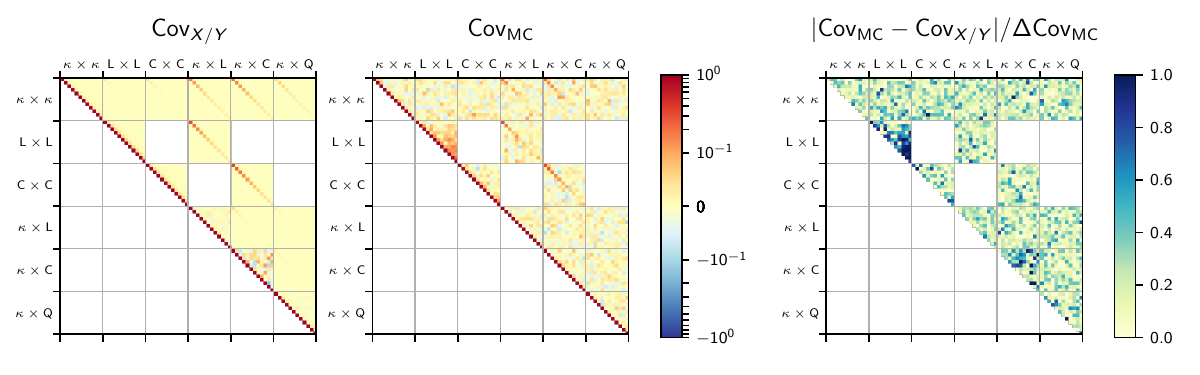}
	\caption{Validation of the covariance matrix. On the left panel, the covariance matrix used in the analysis from eq.~\eqref{eq:covariance}, denoted $\mathrm{Cov}_{X/Y}$; on the middle panel, the empirical covariance matrix of 1000 simulated stacked pseudo spectra, denoted $\mathrm{Cov}_{\mathrm{MC}}$. Both have been normalised by the diagonal elements of $\mathrm{Cov}_{X/Y}$, therefore the diagonal is 1 by construction on the left panel, and the fact that it is very close to 1 on the middle panel proves the agreement between the two estimates. Off-diagonal elements are polluted on $\mathrm{Cov}_{\mathrm{MC}}$ by numerical noise (which is one order of magnitude smaller than the diagonal elements and would reduce with more simulations). On the right panel, the absolute difference between the two estimates of the covariance matrix (non-normalised) is shown, element-wise divided by the standard deviation of $\mathrm{Cov}_{\mathrm{MC}}$ (obtained from bootstrapping the simulated pseudo spectra). Note the different colour scale of the right panel: the deviation is at most of order 1~$\sigma$, showing good agreement between our two estimates and validating eq.~\eqref{eq:covariance}.}
	\label{fig:xcor_test_cov}
\end{figure*}

In order to validate the pseudo spectrum estimator and the semi-analytical expression of the covariance matrix given in eq.~\eqref{eq:covariance}, we generate 1000 sets of four correlated full-sky maps with appropriate auto and cross spectra (for $\{\lowz,\cmass,\qso,\kcmb\}$) computed using eq.~\eqref{eq:limber}, using the \texttt{synfast} function of \texttt{Healpy}\footnote{\texttt{synfast} generates independent identically distributed random normal variables and makes linear combinations of these variables to generate Gaussian distributed spherical Fourier coefficients $a_{\ell m}$ with appropriate covariances.} following a procedure similar to \citet{2015ApJ...802...64B,2016PhRvD..94h3517N}.
{These maps are then masked and their pseudo spectra are compared to the analytical expected value from \cref{eq:mean_ps}, as shown in \cref{fig:pcl_validation}. In order to validate the covariance matrix, a similar set of full-sky maps} are then added realistic noise: for each lensing convergence map, we add an uncorrelated Gaussian noise with spectrum $N^{\kappa}_{\ell}$ given by the approximate spectrum delivered by the Planck Legacy Archive, which is precise enough for the covariance validation. For each galaxy density map, we also need to simulate Poisson sampling. To do so, we generate a map where the value in pixel $p$ is a Poisson random variable of mean $\lambda_p$, i.e.,
\begin{equation}
    n_p \sim \mathrm{Poisson}\left(\lambda_p\right)
    \quad \text{with} \quad
    \lambda_p = \overline{N} \left( 1 + \delta_p \right),
\end{equation}
where $\delta_p$ is the simulated overdensity at pixel $p$ and $\overline{N}$ is the mean number of galaxies per pixel (different for the three samples). A reconstructed density map is then built using eq.~\eqref{eq:density}, which now incorporates Poisson shot-noise. These full-sky maps are then masked and their pseudo spectra are evaluated.
The empirical covariance of the sets of pseudo spectra is finally computed and compared to the semi-analytical covariance we use throughout this analysis. The result in \cref{fig:xcor_test_cov} shows good agreement and validates the estimator and the simulation of the matrices $\mathbfss{X}$ and $\mathbfss{Y}$ (that were computed using generic spectra, see \cref{sec:XY_matrices_in_the_covariance}).

\begin{figure}
	\includegraphics[width=\columnwidth]{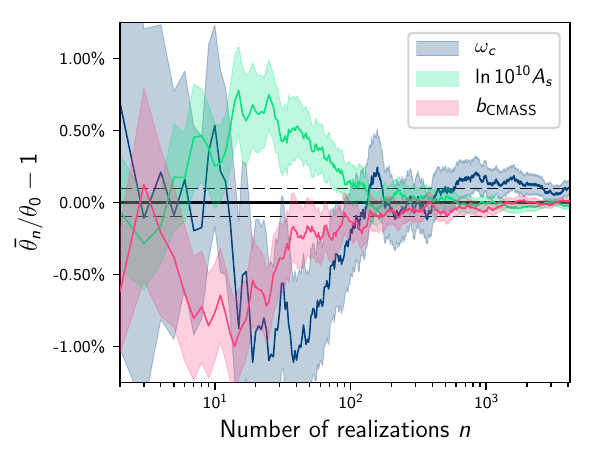}
	\caption{Statistical pipeline validation with Monte-Carlo simulation: realisations of the pseudo spectra are drawn from the likelihood $\mathcal{L}$ and best-fit parameters ${\bmath{\theta} = (\omegac, \ln 10^{10} \As, b_{\cmass} )}$ are computed. The relative error on the mean values of best-fit parameters $\overline{\bmath{\theta}}_n$ as a function of the number of realisations $n$ is shown here. The dotted lines show the 0.1\% requirement for this test, reached after 4146 realisations. The variance (displayed by the coloured bands) decreases as $1/\sqrt{n}$ while the mean values converge towards their input values, demonstrating the internal consistency of the statistical pipeline. Note however the very small deviation on $\omegac$, within the error requirement, but in accordance with the fact that the maximum likelihood estimator is consistent only asymptotically unbiased.}
	\label{fig:mc_test}
\end{figure}

\begin{figure}
	\includegraphics[width=\columnwidth]{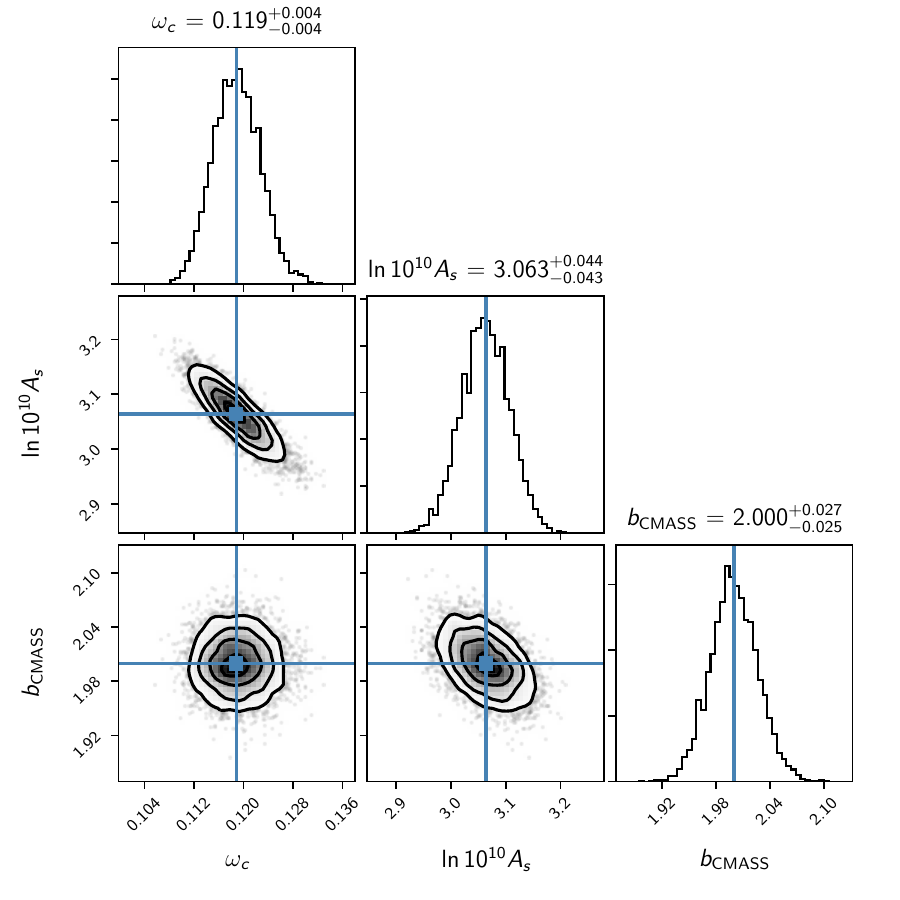}
	\caption{Same test as \cref{fig:mc_test}, now showing the distribution of best-fit parameters for 4146 realisations. The ellipses show the 0.5, 1, 1.5 and 2$\sigma$ contours and the blue lines show the input values of the parameters.}
	\label{fig:mc_test_corner}
\end{figure}

The statistical pipeline is validated by performing a Monte Carlo analysis similar to the one performed in \citet{2014JCAP...05..039P}. Specifically, we want to check if the estimated parameters are unbiased. For that purpose, given the adopted fiducial model, we use the likelihood as the probability distribution of the pseudo spectra $\tilde{C}_{\ell}^{AB}$ to generate random samples (\emph{i.e.} sets of stacked vectors $\tilde{\bmath{C}}^{\mathrm{obs}}$). For each sample, we fit all parameters to be tested, thus building a collection $\{\bmath{\theta}_i\}$ of best-fit values for these parameters. At step $n$, the means $\overline{\bmath{\theta}}_n=\sum_{i=1}^n \bmath {\theta}_i / n$ and standard deviations of the collection of best-fit values are computed. The largest relative error (LRE) over parameter means is computed and we repeat the process, adding more samples, until the LRE has reached a level of 0.1\% and check that the fiducial values are within the error bars. For this test, we only use one sample of galaxies with the redshift distribution of the {\cmass} sample and generate samples of $\tilde{\bmath{C}}_{\ell}^{\kappa_{\mathrm{CMB}} \times \kappa_{\mathrm{CMB}}}$, $\tilde{\bmath{C}}_{\ell}^{\kappa_{\mathrm{CMB}} \times \delta_{\cmass}}$ and $\tilde{\bmath{C}}_{\ell}^{\delta_{\cmass} \times \delta_{\cmass}}$. Results of this test are plotted in \cref{fig:mc_test,fig:mc_test_corner}, showing respectively the evolution of the mean values of the best fit parameters $\overline{\bmath{\theta}}_n$ as a function of the number of realisations $n$ and the distribution of the best-fit parameters for those same realisations. They confirm that the parameters' estimators are unbiased at least at the 0.1\% level.

\subsection{Null tests}
\label{sub:null_tests}

We present in this section null tests that were performed to exclude potential systematic errors related to the masks and selection of the spectroscopic tracers.

\begin{figure}
	\includegraphics[width=\columnwidth]{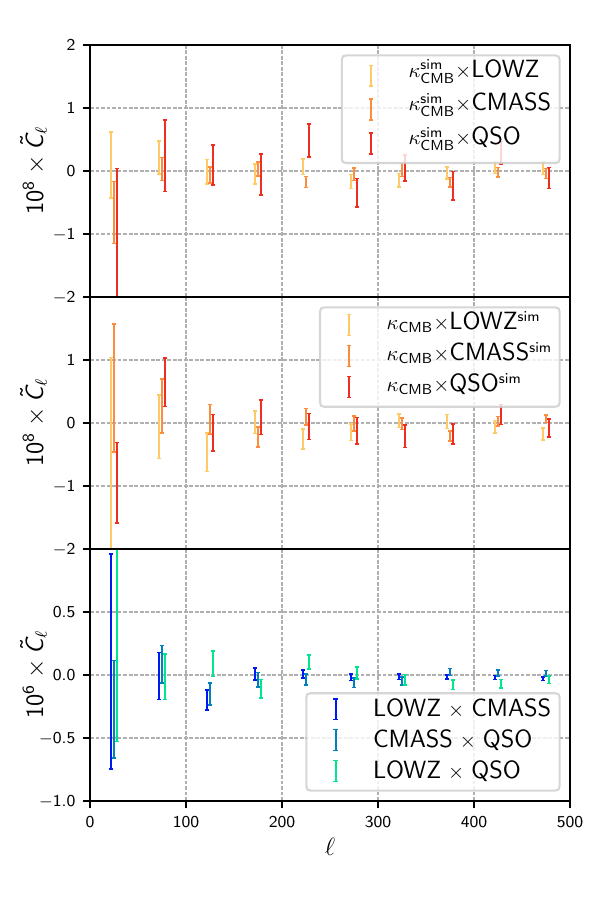}
	\caption{Null test for residual correlation. The top panel shows the mean cross-pseudo spectrum between simulated lensing maps and the real galaxy/quasar density maps, the middle panel shows the mean cross-pseudo spectrum between the real lensing map and simulated galaxy/quasar density maps, and the bottom panel shows the cross-pseudo spectra between the tracers. The cross-correlations with simulated maps are consistent with zero, showing no leakage of power from the masks, while the cross-correlation of the tracers density show marginal correlation, at worst one order of magnitude lower than the autocorrelation signals.}
	\label{fig:null}
\end{figure}

In order to assess potential leakage of power in the cross spectra due to the masking, we cross-correlate the 100 simulated reconstructed lensing maps of the Planck Legacy Archive with the observed density maps of the three galaxy samples, and then correlate the observed lensing map with 100 simulated galaxy maps. This procedure removes cosmological angular correlation, and what correlation remains will be linked to the masks themselves. We find that all results are consistent with no correlation, excluding strong contamination from masking. We also measure the cross spectra between the galaxy and quasars sample and find marginal correlations, well below the auto-correlation signals.

\begin{figure}
    \includegraphics[width=\columnwidth]{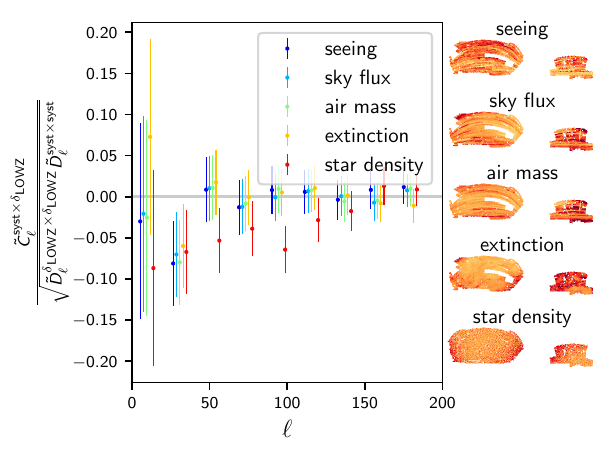}
    \caption{{Null test of photometric systematics contamination: normalised large-scale cross spectra of photometry-related measurements with the density of \textsf{LOWZ} galaxies. They are statistically consistent with zero, with a small anti-correlation with stellar density. For the seeing, sky flux and extinction, we repeated the measurement in the $g$, $r$ and $i$ bands and found very similar results (only the $g$ band measurement is shown for clarity). The maps of these observables are shown on the right in equatorial coordinates where the north and south galactic caps of the SDSS survey can be seen. The $\chi^2$ statistics for these cross spectra are respectively 215, 192, 204, 216 and 235 for 192 degrees of freedom ($0<\ell\leq192$) for the systematics maps shown on the right, excluding a large contamination.}
    }
    \label{fig:LOWZ_phot_syst}
\end{figure}

\begin{figure}
    \includegraphics[width=\columnwidth]{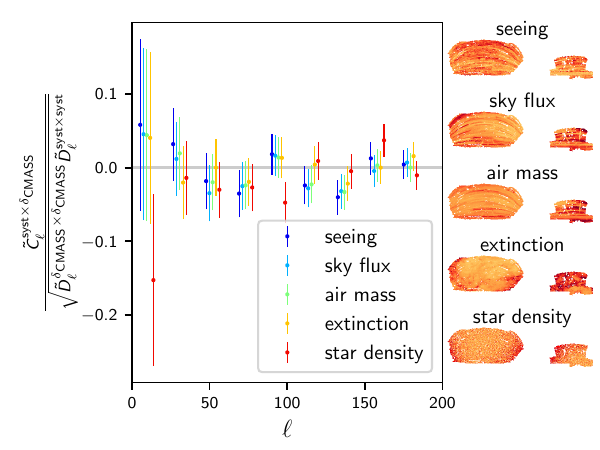}
    \caption{Same as \cref{fig:LOWZ_phot_syst} for the \textsf{CMASS} galaxies. The $\chi^2$ statistics are respectively 218, 178, 204, 194 and 251.}
    \label{fig:CMASS_phot_syst}
\end{figure}

\begin{figure}
    \includegraphics[width=\columnwidth]{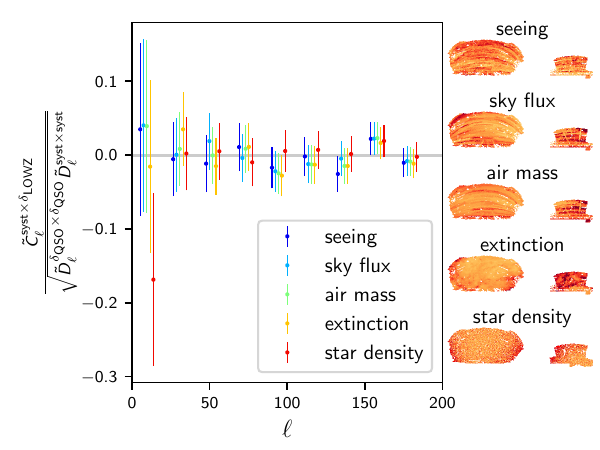}
    \caption{Same as \cref{fig:LOWZ_phot_syst} for the quasar sample. The $\chi^2$ statistics are respectively 207, 211, 227, 241 and 295.}
    \label{fig:QSO_phot_syst}
\end{figure}

Variable observational conditions during the SDSS photometric survey could potentially result in non-uniform selection functions of the galaxy and quasar samples, and introduce artificial power in the auto spectra at large scales. In order to exclude dramatic power leakage, we constructed maps of resolution {$N_{\mathrm{side}}=128$ of the seeing, sky flux, extinction (for the $g$, $r$ and $i$ bands) and air mass of the photometric observations} that were used to select galaxies and quasars in the spectroscopic catalogs,\footnote{To do so, we made use of the CasJob service of the SDSS SkyServer, at \url{http://skyserver.sdss.org/casjobs/}.} {as well as a map of the stellar density using the same cuts as in \citet{2017MNRAS.464.1168R}}. We then verify that the cross power spectra with the density maps (built in \cref{ssub:building_the_maps}) are consistent with a null value. To do so, we {use the covariance matrix of the pseudo cross spectra given in \cref{eq:covariance}, taking advantage of the fact that the masks are the same, such that the covariance matrix of reduced pseudo cross spectra, defined as
\begin{equation}
    \tilde{\rho}_\ell \equiv \frac{\tilde{C}_{\ell}^{\mathrm{syst} \times \delta_{g}}}{\sqrt{\tilde{D}_{\ell}^{\delta_{g} \times \delta_{g}} \tilde{D}_{\ell}^{\mathrm{syst} \times \mathrm{syst}}}},
    \label{eq:rho_ell}
\end{equation}
is {approximated} by
\begin{equation}
    \mathrm{Cov}\left( \tilde{\rho}_\ell, \tilde{\rho}_\ell \right) \approx \frac{1}{f_{\rm sky}^2 (2\ell'+1)} M_{\ell \ell'}^{gg}
    \label{eq:cov_rho_ell}
\end{equation}
where $M_{\ell \ell'}^{gg}$ is the mixing matrix associated with the masks of the \lowz, \cmass and \qso samples, and $\tilde{D}_\ell$ denotes the pseudo auto power spectrum including noise. Here we have approximated the ratio of full-sky to partial sky spectra as $\tilde{D}_{\ell}^{AA} \approx f_{\rm sky}^A D_{\ell}^{AA}$ {and have neglected the variance\footnote{{Note that we consider here the covariance over cosmological realizations at fixed realization of the observational systematics.}} of $\sqrt{\tilde{D}_{\ell}^{\delta_{g} \times \delta_{g}}}$ in the denominator in \cref{eq:rho_ell} since it is measured at high signal-to-noise ratio (which, because of cosmic variance, is a ${\mathcal{O}( 1 / \sqrt{\ell})}$ approximation further tamed by the square root)}.
}
{\Cref{fig:LOWZ_phot_syst,fig:CMASS_phot_syst,fig:QSO_phot_syst} show the measured cross power spectra $\tilde{C}_{\ell}^{\mathrm{syst} \times \delta_{g}}$ for multipoles ${0 < \ell \leq 192}$. $\chi^2$ statistics are measured (values are given in the legends of the figures) and consistent with no correlation\footnote{The stellar density shows significant anti-correlation with the quasar density, but only for the smallest multipoles which we discard. Above $\ell>20$, it is consistent with zero.}, with no discernable trend across multipoles.}
From these tests, we conclude that photometric systematics do not strongly correlate with the overdensity maps.

\section{Analyses and results}
\label{sec:analyses_and_results}

\subsection{Cosmological model}
\label{sub:cosmological_model}

Our base model is the standard \LCDM model with flat spatial sections (hence {$\Omega_k=0$}) and a DE component with equation of state $w=-1$. The base parameters are the present-day baryon and cold dark matter densities, $\omegab \equiv \Omegab h^2$ and $\omegac \equiv \Omegac h^2$, respectively -- where $\Omega_i = \rho_i / \rho_c$ is the ratio of the component's energy density to the critical energy density $\rho_c$--, the Hubble constant today $H_0 = \SI{100}{\h\km\per\s\per\mega\parsec}$, the redshift of reionization $\zre$, the logarithm of the primordial curvature $\zeta$ dimensionless power spectrum $\ln 10^{10}\As$ and its tilt $\ns$ such that
\begin{equation}
	\mathcal{P}_\zeta (k) = \As \left( \frac{k}{k_0} \right)^{\ns-1},
\end{equation}
with the pivot scale $k_0 = \SI{0.05}{\per\mega\parsec}$. We include massive neutrinos, parametrised by the effective number of neutrinos in the relativistic limit $N_{\rm eff}=\num{3.046}$ (taking into account non-instantaneous decoupling), an effective temperature $T_{\nu}/T_{\gamma}=\num{0.71611}$, where $T_{\gamma}$ is the photon temperature \citep[slightly departing from $\left(4/11\right)^{1/3}$ to take into account neutrino heating from electron/positron annihilation, see][]{Lesgourgues:2009hi}, and using one massive neutrino of mass ${\mnu=\SI{0.06}{\eV}}$ and two massless neutrinos, consistent with the Planck base \LCDM model. The linear matter power spectrum $P_\mathrm{m}(k,z)$ and the CMB temperature power spectrum $C_\ell^{TT}$ are computed using the Cosmic Linear Anisotropy Solving System (\texttt{CLASS}) as a backend to \texttt{NumCosmo}. The non-linear matter power spectrum is computed using a \texttt{halofit} prescription \citep{2003MNRAS.341.1311S} implemented in \texttt{NumCosmo}, with parameters from \citet{2012ApJ...761..152T}, modified to take into account neutrinos as in \texttt{CLASS}.
Reionization is modelled in a \texttt{CAMB}-like fashion \citep{2000ApJ...538..473L} and parametrised by the mid-point $\zre$, fixed width $\Delta \zre = 0.5$, and includes Helium reionization at a fixed redshift $z_\mathrm{re}^\mathrm{He}=3.5$. Recombination is computed within \texttt{CLASS} and Big Bang nucleosynthesis is computed with \texttt{PArthENoPE}\footnote{\url{http://parthenope.na.infn.it/}}\citep{2008CoPhC.178..956P}.

From the constraints that we will obtain in our analyses, we will also estimate the total matter density parameter\footnote{The DE density parameter today is $\Omega_{\Lambda} \approx 1-\Omegam$ since we consider only flat space sections (neglecting radiation).} ${\Omegam = \Omegab + \Omegac + \Omega_{\nu}}$, the optical depth to the last scattering surface $\tau$ and the variance of the linear matter density fluctuations $\sigma_8^2$ in spheres of radius $R = \SI{8}{\per\h\mega\parsec}$ extrapolated to $z=0$,
\begin{equation}
	\sigma_8^2 = \int dk \frac{k^2}{2\pi^2} P_\mathrm{m}(k,z=0) \left|W(k,R)\right|^2,
\end{equation}
where the top-hat window function is $W(k,R) = 3 j_1(kR)/kR$ and the matter power spectrum is computed from linear theory.

\subsection{Statistical analysis}
\label{sub:statistical_analysis}

In this section, we describe our Bayesian statistical analysis and present constraints on cosmological parameters and BOSS spectroscopic tracers' biases.

We first apply the Markov Chain Monte Carlo (MCMC) approach using only CMB lensing-LSS correlations data -- the set of the six auto and cross spectra of CMB lensing and BOSS galaxy and quasar overdensities, as shown in \cref{fig:clobs} -- and varying only a subset of cosmological parameters in order to assess the constraining power of these. In particular, we also consider different combinations of the auto and cross spectra to measure the effects on the parameter constraints provided by these probes.
Then, we add CMB temperature information and obtain constraints on the \LCDM model and extensions including the {total} mass of neutrinos $\Smnu$ -- that impacts small-scale structure formation -- and the DE equation of state $w$ -- that impacts the expansion in the low redshift Universe.

We perform MCMC analyses using an \emph{ensemble sampler}\footnote{Some authors refers to ensemble samplers as \emph{population Monte Carlo}.} with many walkers {(32 to 1000)}, moving their positions in the parameter space as an ensemble \emph{via} a \emph{stretch move} scheme \citep{Goodman:2010et} implemented in \texttt{NumCosmo}. We monitored the convergence of the chains using three numerical tools, namely the Multivariate Potential Scale Reduction Factor~\citep[MPSRF,][]{1992StaSc...7..457G, Brooks:2012ju}, the Heidelberger-Welch test \citep{Heidelberger:1981ih, Heidelberger:1983ia} and the Effective Sample Size (ESS); see \cref{sec:mcmc_convergence_tests} for more details.

These diagnostics can fail in different situations. For this reason, we also performed three different visual inspections for each parameter:
\begin{enumerate}
    \item the parameter trace plot, \emph{i.e.} the value of the parameter for a given walker \emph{vs} iteration time.
    \item the ensemble distribution trace plot, that is, the empirical ensemble distribution given by the walkers' positions \emph{vs} iteration time. This allows us to monitor the evolution of the ensemble mean and variance.
    \item the total mean \emph{vs} the cumulative sum of the ensemble means: if the chain has reached convergence, the difference (scaled by the spectral density at null frequency) is distributed as a brownian bridge, the $L_2$ norm of which is used in the Schruben test.
\end{enumerate}

For all MCMCs, we ran them until all the relative errors of the means were smaller than $10^{-2}$; at this point, we applied all the
tests above, and if the chains failed some of them, we continued the run
until all tests were satisfied.

\subsubsection{Constraints on $\sigma_8$ and $\Omegam$ from CMB lensing--LSS correlations {only}} 
\label{ssub:constraints_from_cmb_lensing_large_scale_structure_correlations}

\begin{figure*}
\centering
	\includegraphics[width=17cm]{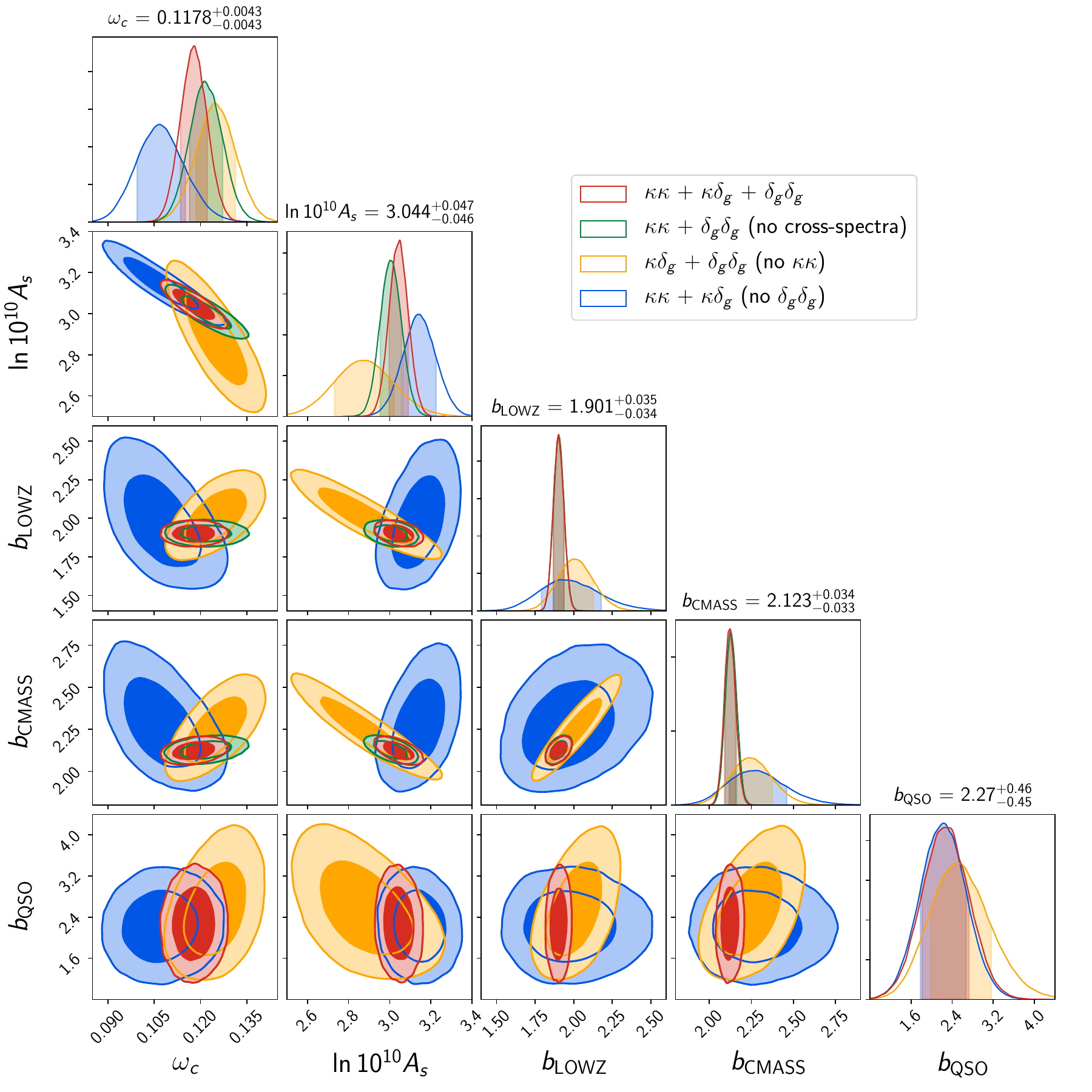}
	\caption{
    {Posterior distribution on a selection of cosmological parameters ($\omegac$ and $\As$) and tracer clustering biases from the ``CMB lensing$\otimes$LSS'' data set (see \cref{fig:clobs}). Other cosmological parameters ($H_0$, $\omegab$, $\ns$ and $\zre$) are fixed at Planck 2015 ``TT,TE,EE+lowP+lensing+ext'' best fit values. The two-dimensional projections show the 68\% and 95\% confidence levels. The (${\ln 10^{10} \As}$,$\omegac$) plane illustrates the degeneracy breaking and the confidence region shrinkage due to the addition of the cross spectra (compare the red and green contours). The integral of the histograms are normalised to unity, therefore in the approximation of Gaussian distributions, the maxima of the histograms are inversely proportional to the standard deviation of the parameters, allowing to directly read the improvement of the constraints. Note that the quasar bias is not fitted for the subset including only auto spectra (``{$\kappa\kappa + \delta_g \delta_g$}'').}
    }
	\label{fig:mcmc_xcor_only}
\end{figure*}

Data from CMB lensing and spectroscopic tracers of matter alone cannot efficiently constrain all cosmological parameters. However, we want to highlight the cosmological information carried by these probes. To do so, we perform several MCMC analyses considering
a subset of free cosmological parameters. These are only illustrative in the sense that the posterior distribution of cosmological parameters will be shrunk from fixing some others. 

The theoretical spectra have different dependences on the cosmological parameters and galaxy/quasar biases. The most explicit dependencies of the angular spectra $\tilde{C}_{\ell}$ are on the power spectrum amplitude ($\As$ or $\sigma_8$), the matter density parameter $\Omegam$ and the galaxy/quasar biases (see the kernels in \cref{sub:cross-correlation}):
\begin{align}
	C_{\ell}^{\kappa \kappa} & \propto  \Omegam^2 \As \\
	C_{\ell}^{\kappa \delta_g} & \propto  \Omegam b_g \As \\
	C_{\ell}^{\delta_g \delta_g} & \propto  b_g^2 \As.
\end{align}
This system is closed, \emph{i.e.} in principle, comparing the various auto and cross spectra should allow for an non-degenerate estimation of the parameters.

Therefore, we run MCMCs freeing $\omegac$, $\ln 10^{10} \As$ and the galaxy/quasar biases, and fixing all other cosmological parameters. Their fiducial values are from Planck 2015 ``TT,TE,EE+lowP+lensing+ext'' best fits \citep{2016A&A...594A..13P}. We assume flat prior distributions over wide ranges (larger than the sampled ranges). In order to distinguish and quantify the information contained in the various measured auto and cross spectra, we try different combinations. That is, we run an MCMC with the full dataset (``$\kappa\kappa$+$\kappa$$\delta_g$+$\delta_g$$\delta_g$''), and then repeat without the cross spectra (``$\kappa\kappa$+$\delta_g$$\delta_g$''), without the CMB lensing auto spectrum (``$\kappa$$\delta_g$+$\delta_g$$\delta_g$'') and without the galaxy auto spectra (``$\kappa\kappa$+$\kappa$$\delta_g$''). We run these chains with 100 walkers to ensure a good mixing. Their MPSRFs are below 1.02 and the correlation lengths are of order 20-40, varying amongst parameters.

\begin{figure}
	\includegraphics[width=\columnwidth]{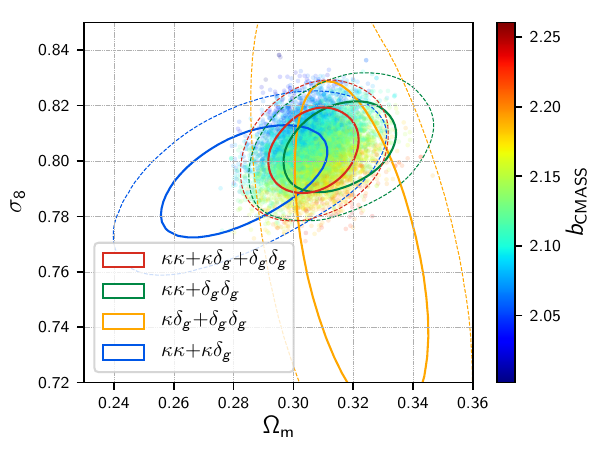}
	\caption{Confidence regions for $\sigma_8$ and $\Omegam$ corresponding to the distributions sampled by the ensemble sampler MCMC algorithm for the CMB lensing-LSS correlations dataset only, with the other cosmological parameters fixed at Planck 2015 ``TT,TE,EE+lowP+lensing+ext'', and various subsets of the dataset. The inner solid (outer dashed) contours give the 68\% (95\%) confidence levels. CMB lensing and galaxy densities show different degeneracies, that are partially broken by combining the observations. Coloured points are samples from the full dataset chains, that show how $\sigma_8$, $\Omegam$ and galaxy biases are degenerated.}
	\label{fig:sigma8_Omegam}
\end{figure}

The sampled posterior distributions of these parameters are shown in \cref{fig:mcmc_xcor_only} for the full dataset and the three subsets aforementioned.
We note that the ``$\kappa$$\delta_g$+$\delta_g$$\delta_g$'' and ``$\kappa\kappa$+$\kappa$$\delta_g$'' subsets, dominated by respectively galaxy clustering and CMB lensing information, provide complementary information, since the correlations between the parameters, except those in the ($b_\textsf{LOWZ}$, $b_\textsf{CMASS}$) plane, present different alignments (see the blue and yellow confidence regions in \cref{fig:mcmc_xcor_only}). Therefore, the constraints on the parameters are greatly improved when combining both auto spectra, ``$\kappa\kappa$+$\delta_g$$\delta_g$''. Apart from the constraints on $b_\textsf{LOWZ}$, $b_\textsf{CMASS}$, which are already strongly determined by galaxy density auto spectra, the additional information contained in the cross spectra narrows the distribution, as can be observed in the (${\ln 10^{10} \As}$,$\omegac$) plane by comparing the confidence regions for ``$\kappa\kappa$+$\delta_g$$\delta_g$'' (in green) with ``$\kappa\kappa$+$\kappa$$\delta_g$+$\delta_g$$\delta_g$'' (in red). The addition of the cross spectra decreases the statistical error by 10\% for ${\ln 10^{10} \As}$ and 20\% for $\omegac$, and slightly shifts the best fits (by less than {1~$\sigma$}). This plane is translated into the ($\sigma_8$,\;$\Omegam$) plane in \cref{fig:sigma8_Omegam} where the degeneracy breaking expected from the joint analysis is highlighted.

\subsubsection{Cosmological constraints from the joint analysis of Planck and BOSS data} 
\label{ssub:cosmological_constraints_from_the_joint_analysis_of_planck_and_boss_data}



In this section, we carry out the analysis combining CMB temperature and the joint likelihood of CMB lensing and
galaxy/quasar densities used in the previous section (that is all six power spectra of \cref{fig:clobs})
to {derive cosmological constraints. First, we constrain the 6-parameters \LCDM model (with parameters $\omegab$, $\omegac$, $H_0$, $\zre$, $\ln 10^{10} \As$ and $\ns$) and compare constraints from the full joint analysis (``Planck $TT$ + lensing $\otimes$ BOSS tracers'') with that derived from CMB temperature anisotropies only (``Planck $TT$'') or CMB temperature and lensing (``Planck $TT$ + lensing''). For the joint analysis, we will also consider two cuts as mentioned in \cref{sub:cross_correlation_data}.}
Note that we neglect the correlation between CMB temperature and the matter density at later times
(either baryonic matter in galaxies and quasars or dark matter weighted by CMB lensing), \emph{i.e.}, we neglect the late
ISW effect, as it is not yet detected with a strong significance
\citep{2016A&A...594A..21P,2016PhRvD..94h3517N}, and we discuss possible consequences in \cref{sub:discussion}. In practice, this means that we approximate the total likelihood by the
product of the CMB temperature and CMB lensing-LSS correlations likelihood functions.

We use the Planck likelihood codes \texttt{Plik} and \texttt{Commander} \citep{2016A&A...594A..11P} respectively
 for high and low multipoles of the temperature-only power spectrum $C_{\ell}^{TT}$. The likelihood codes
introduce 15 additional nuisance parameters related to foreground and instrument models \citep[$A^{\mathrm{CIB}}_{217}$,
$\xi^{\mathrm{tSZ}\times \mathrm{CIB}}$, $A^{\mathrm{tSZ}}$, $A^{\mathrm{PS}}_{100}$, $A^{\mathrm{PS}}_{143}$,
$A^{\mathrm{PS}}_{143\times 217}$, $A^{\mathrm{PS}}_{217}$, $A^{\mathrm{kSZ}}$, $A^{\mathrm{dust}TT}_{100}$,
$A^{\mathrm{dust}TT}_{143}$, $A^{\mathrm{dust}TT}_{143 \times 217}$, $A^{\mathrm{dust}TT}_{217}$, $c_{100}$, $c_{217}$ and $y_{\mathrm{cal}}$; see][]{2016A&A...594A..11P}. We use the profile likelihood to speed up our MCMC analyses, subfitting the nuisance parameters for each set of cosmological parameters. We describe this methodology in \cref{sec:profile_likelihood}, and also show that it does not affect the results on cosmological parameters.

{We use flat, large priors on all cosmological parameters and tracer biases, common to all MCMC runs. The lower and upper limits are given in \cref{tab:constraints}, together with constraints derived from the (optimistic) joint analysis.
}

\paragraph{Contraints on \LCDM.} 
\label{par:contraints_on_lcdm}


\begin{figure*}
\centering
	\includegraphics[width=17cm]{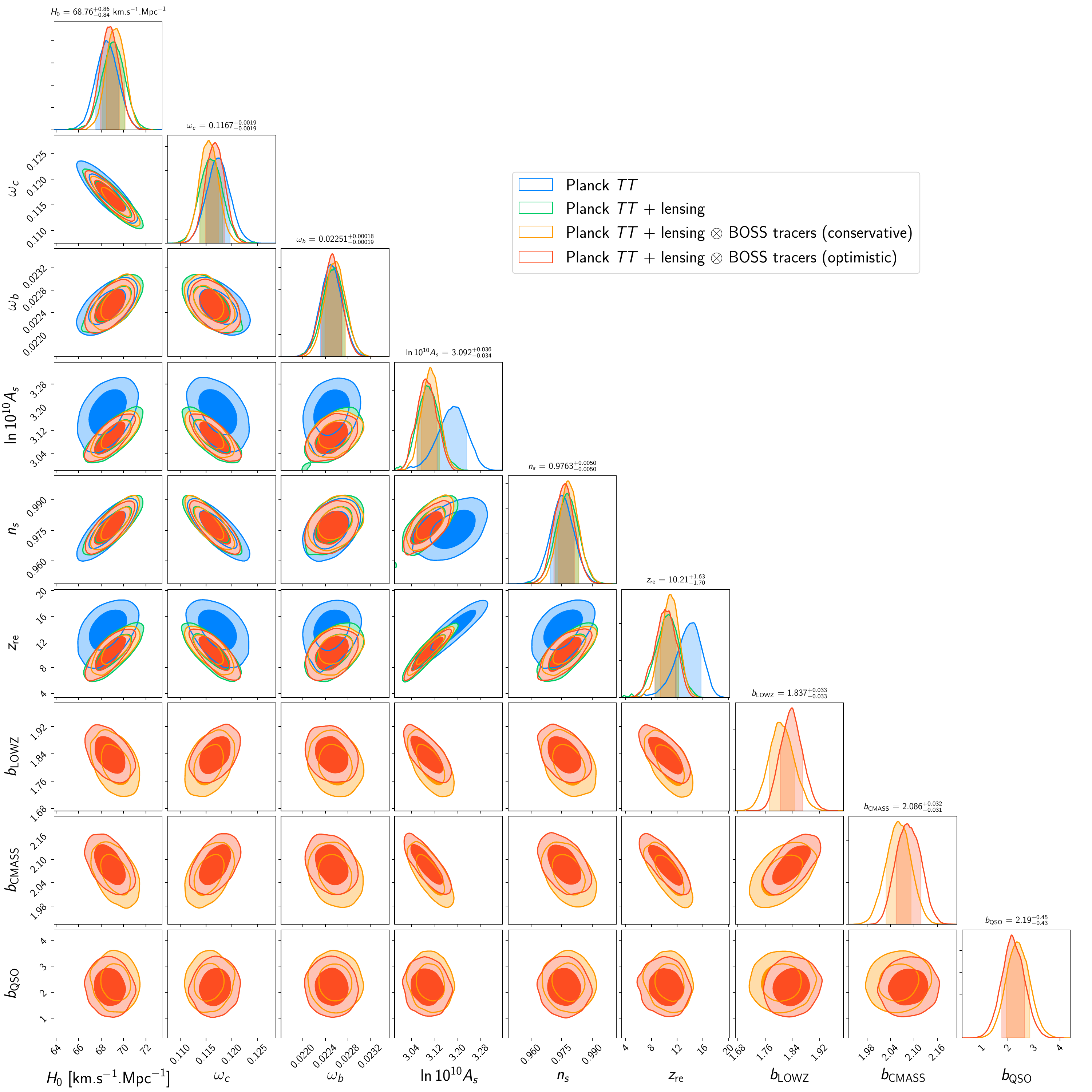}
	\caption{Constraints on the parameters of the base 6-parameters \LCDM model and spectroscopic tracers biases. Confidence regions (68\% and 95\% levels) are shown respectively in blue, green and red for CMB temperature only, CMB temperature combined with CMB lensing, and the joint analysis of CMB temperature and the correlations of CMB lensing and LSS tracers. The constraints above the marginal posteriors are for this last data set.}
	\label{fig:LCDM6params_corner}
\end{figure*}

\begin{figure}
	\includegraphics[width=\columnwidth]{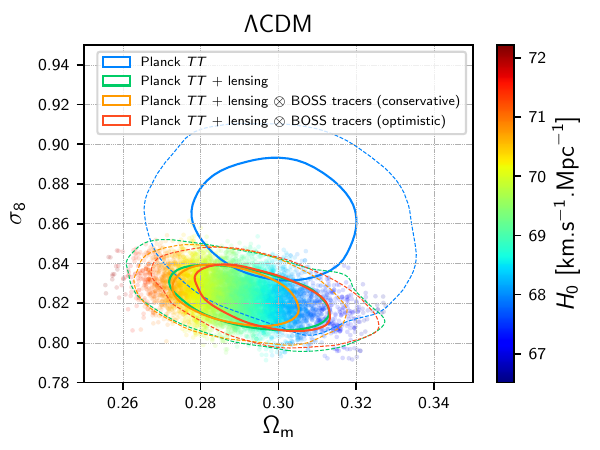}
	\caption{Constraints on $\sigma_8$ and $\Omegam$ for the 6-parameter base \LCDM model from the combination of CMB temperature and the correlations of CMB lensing and spectroscopic tracers. The 1 and 2~$\sigma$ contours are represented by the respectively solid and dashed lines. The coloured points show the degeneracy with $H_0$ and are samples from the ``{CMB $TT$ + CMB lensing $\otimes$ LSS}'' chain.}
	\label{fig:xcorfull_PlanckTT_LCDM_sigma8_Omegam}
\end{figure}

\Cref{fig:LCDM6params_corner} shows the constraints on the base \LCDM model's parameters for the three aforementioned data combinations. When using CMB $TT$ only, we find parameter constraints that are in perfect agreement with the Planck analysis\footnote{The detailed results of the Planck MCMC analyses are available here: \url{https://wiki.cosmos.esa.int/planckpla2015/images/f/f7/Baseline_params_table_2015_limit68.pdf}.}.
The strong degeneracy observed between the power spectrum amplitude $\As$ and the reionization redshift $\zre$ corresponds to the amplitude of the power spectrum of CMB temperature anisotropies, which is proportional to $\As e^{-2\tau}$ where $\tau$ is the optical depth to the last scattering surface, strongly dependent on reionization history. Adding CMB lensing drives $\As$ and $\zre$ towards lower values along this degeneracy with a shift of about $1\sigma$ for each parameter
as evinced by the one- and two-dimensional projection of the posterior distributions.
Finally, adding information of LSS tracers (both the auto-correlations and cross-correlations with CMB lensing) provides only slightly smaller contours for these parameters and, therefore,
they do not significantly help in breaking this degeneracy. We find no significant improvement for $\tau$ or $\sigma_8$, although it is consistent with the constraints from CMB lensing. In the ($\sigma_8,\;\Omegam$) plane (see \cref{fig:xcorfull_PlanckTT_LCDM_sigma8_Omegam}), we observe that early Universe data favors bigger values of $\sigma_8$ than the late one,
as repeatedly reported in the literature \citep{2017MNRAS.465.1454H,2016A&A...594A..24P}. Whether this is indication of new physics or a systematics artefact is beyond the scope of this work, but it might be an important issue in the future.

Nonetheless, there is an improvement of order 20\% on the measurements of $H_0$ and $\omegac$ for both the optimistic and conservative cuts, {and the total volume in parameter space is notably reduced. This can be quantified in the approximation of Gaussian posteriors by computing a figure of merit ${\mathrm{FoM}\equiv\abs{\mathrm{Cov}(\Theta_{\rm cosmo})}^{-1/2}}$, where $\mathrm{Cov}(\Theta_{\rm cosmo})$ is the empirical covariance matrix of cosmological parameters. Relative to the CMB $TT$ only posterior, adding CMB lensing increases the FoM by a factor 2.0, and the full joint analysis increases the FoM by a factor \num{2.9}.} We find a best-fit of {${H_0 = \SI{68.8}{\km\per\s\per\mega\parsec}}$}, slightly higher than CMB temperature alone -- albeit still lower than distance measurements from {supernov\ae} \citep{2016ApJ...826...56R} or time delays in strong lensing \citep{2017MNRAS.465.4914B} -- and ${\omegac = 0.117 \pm 0.002}$ as the degeneracy between $\As$ and $\omegac$ is broken by the lensing--LSS correlations (see \cref{fig:mcmc_xcor_only}). This results in a constraint on the matter density parameter ${\Omegam = 0.296 \pm 0.011}$. Additionally, we obtain strong constraints on the biases of the galaxy samples, respectively
{
\begin{equation}
    \begin{array}{ll}
        b_{\lowz} &= 1.837 \pm 0.033 \\
        b_{\cmass} &= 2.086 \pm 0.032. \\
    \end{array}
\end{equation}}
{These 4\% constraints are in general agreement with previous measurements using angular power spectra \citep{2012ApJ...761...14H}), which have the advantage of being model-independent in the sense that estimating angular power spectra $\tilde{C}_{\ell}$ does not require any assumption on cosmology since we don't measure distances}. Moreover, all the cosmological parameters of the \LCDM model are fitted alongside. Note, however, that our modelling assumes a constant bias, that can be interpreted as a redshift- and scale-averaged bias, when other analyses exploring the non-linear regime used a scale-dependent bias (in the form of a Taylor expansion, \emph{e.g.} in \citet{2017MNRAS.465.1757G}) or simply more redshift bins. Interestingly, the analysis also shows significant correlations between the biases and cosmological parameters, in particular with $\omegac$, $H_0$ and $\As$. If one considers biases as effective parameters encoding structure formation and clustering of galaxies, these correlations can shed light on the astrophysical and cosmological processes governing the formation of such structures. Finally, we also obtain a broad constraint on the bias of the uniform sample of quasars from the cross-correlation with CMB lensing
{
\begin{equation}
    \begin{array}{ll}
        b_{\qso} & = 2.20 \pm 0.44.
    \end{array}
\end{equation}
}
This value is in tension with other measurements \citep{2012MNRAS.424..933W,2015MNRAS.446.3492D,2016JCAP...11..060L,2017JCAP...07..017L} that found a bias of order 3 to 4 (although they did assume a cosmology){, but in agreement with \citet{Alonso:2017tl} that also uses the cross angular power spectrum.}
We found no difference when fitting for this bias when using data from only the northern or southern galactic caps, excluding a possible strong asymmetry and contamination of higher multipoles. We note a surprising trough in the $\tilde{C}_{\ell}^{\kappa_{\mathrm{CMB}} \times \mathrm{QSO}}$ cross spectrum (see \cref{fig:clobs}) around $\ell \sim 400$ that we could not explain. However, this bias directly depends on the amplitude of lensing that may be underestimated \citep{2014A&A...571A..17P}.

{Finally, we observe that both cuts yield very similar results for cosmological parameters, with little improvement from the optimistic cut for the \LCDM model (which can arise from small tensions between data sets). The conservative cut yields lower clustering biases for \lowz and \cmass galaxies, while remaining within $1\sigma$ of the optimistic cut (that includes more non-linear scales).}


\paragraph{Constraints on the total mass of neutrinos $\Smnu$ and the dark energy equation of state $w$.} 
\label{sub:constraints_on_lambda_mathrm_cdm_mnu_and_w_cdm_from_cmb_temperature_and_cmb_lensing_large_scale_structure_correlations}


\begin{figure}
\centering
	\includegraphics[width=\columnwidth]{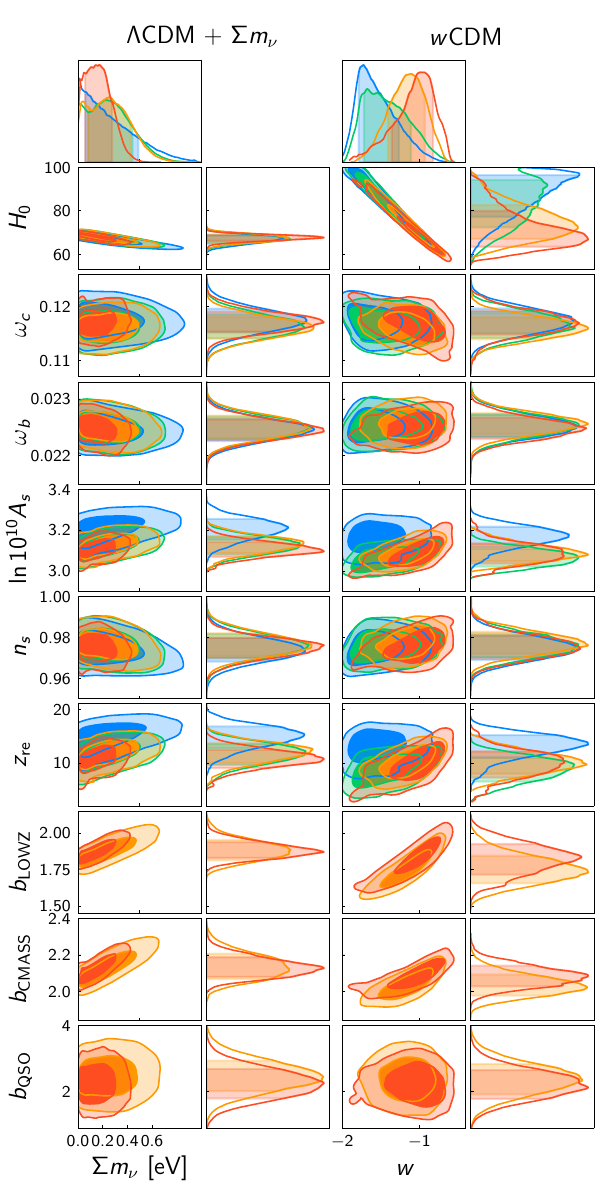}
	\caption{Cosmological constraints on two one-parameter extensions of the base 6-parameters \LCDM model used here. On the column labeled $\Lambda\mathrm{CDM}+\Smnu$, the total mass of the neutrinos $\Smnu$ (expressed in eV) is set free and sampled in addition to the six cosmological parameters and the galaxy and quasar biases, $w$ being fixed to $-1$. On the column labelled $w\mathrm{CDM}$, $w$ is set free and the {total} mass of neutrinos is fixed at its fiducial value of \SI{0.06}{\eV}. Only the two-dimensional distributions involving $\Smnu$ or $w$ are shown, together with marginal posteriors for the other cosmological parameters (rotated to match the leftmost vertical axes). Both columns use the same intervals for comparison. The two upmost plots show the marginal distributions obtained for $\Smnu$ and $w$. Colours are the same as in \cref{fig:LCDM6params_corner}: blue is for CMB temperature alone, green is for CMB temperature and lensing, orange and red are for the joint analysis (conservative and optimistic data cuts).}
	\label{fig:LCDM6params_extensions}
\end{figure}

\begin{figure}
	\includegraphics[width=\columnwidth]{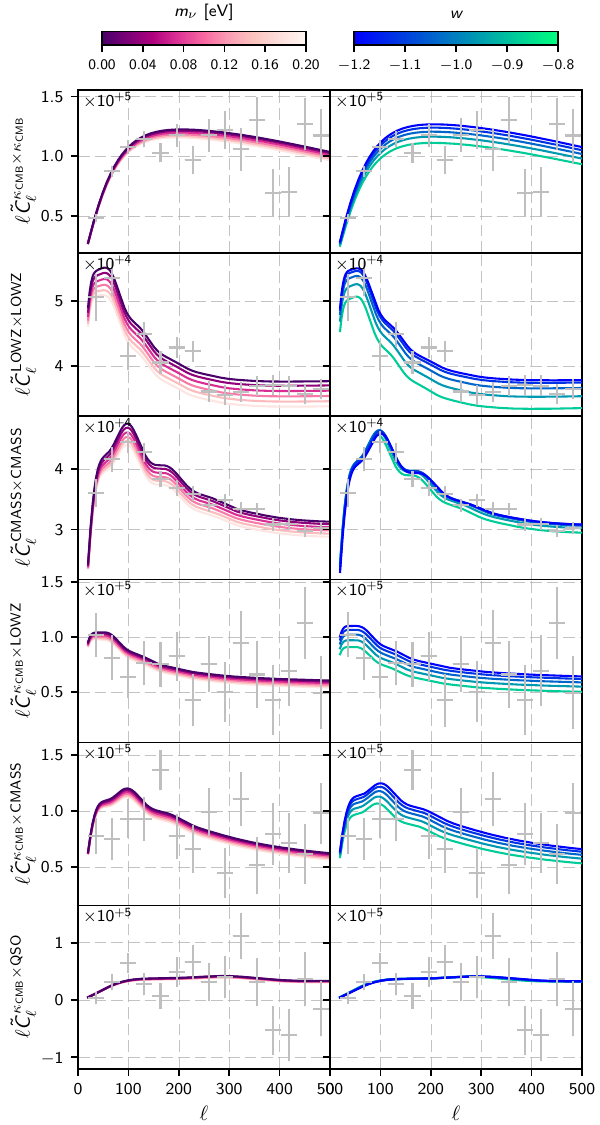}
	\caption{Theoretical pseudo spectra for different values of the total mass of neutrinos $\Smnu$ and $w$. On the left column, $\Smnu$ varies from \SI{0}{\eV} (dark purple) to \SI{0.2}{\eV} (light pink); on the right column, $w$ varies from \num{-1.2} (green) to \num{-0.8} (blue). Data points from \cref{fig:clobs} are overlaid in light grey.}
	\label{fig:dCldparams}
\end{figure}

In the next set of MCMCs, we additionally sample the {total} mass of neutrinos $\Smnu$ (with one massive and two massless neutrinos) or the DE equation of state $w$ (where $w$ is constant over time) separately and compare the performance of the joint analysis in these extended models.

In the first case, we find that the joint analysis yields an upper bound on the total mass of neutrinos of {${\Smnu < \SI{0.28}{\eV}\ [68\%]}$ ($<\SI{0.39}{\eV}\ [95\%]$)} {with the optimistic} cut, dividing the higher bound by a factor of two with respect to the constraint from CMB $TT$ alone (see \cref{fig:LCDM6params_extensions}). We do not detect a total neutrino mass significantly different from zero, but the best fit we obtain around {$m\sim\SI{0.15}{\eV}$} is in agreement with lower bounds around \SI{0.05}{\eV} derived from neutrino oscillations \citep{Olive:1753419}, and in agreement with cosmological upper bounds around \SI{0.12}{\eV}, \emph{e.g.} that derived from the combination of CMB, either with the Lyman-$\alpha$ forest power spectrum \citep{2015JCAP...02..045P} or with BAO measurement \citep{2017PhRvD..96l3503V}. However, the joint analysis with the conservative cut does not improve the upper bound when compared to the ``Planck $TT$ + lensing'' case. As shown in \cref{fig:dCldparams}, the galaxy auto power spectra are sensitive to the total mass of neutrinos as they can probe relatively small scales at low redshift, where massive neutrinos tend to smooth out density fluctuations. {Therefore, the improved constraints with the optimistic cut is likely to be due to small scale contributions.} Moreover, this also means that the total mass of neutrinos should be positively correlated with galaxy biases, which is indeed observed in the lower panels of \cref{fig:LCDM6params_extensions}. We also observe that adding LSS information significantly improves the constraints on the other cosmological parameters in this extended model{, for both the conservative and optimistic cuts}. Because of the anti-correlation between $H_0$ and the total mass of neutrinos, the joint analysis favours a higher expansion rate {${H_0 = \SI[parse-numbers=false]{67.6_{-1.4}^{+1.3}}{\km\per\s\per\mega\parsec}}$} (optimistic case) than CMB data alone. It also noticeably shifts the posterior distributions for $\zre$ and $\As$ towards lower values, resulting in a lower value of the reionization optical depth {${\tau = 0.088 \pm 0.020}$} (though still higher than CMB polarization).

In the second case, we release $w$, the {sum of the neutrino masses} being fixed to ${\Smnu = \SI{0.06}{\eV}}$. CMB temperature anisotropies are only very weakly sensitive to DE and CMB lensing probes the Universe at redshift $z \sim 2$ where matter is still dominating. Therefore these probes do not contain much information on $w$. {We observe a strong anti-correlation between $w$ and the Hubble parameter $H_0$, meaning that observations can be matched by a more slowly expanding Universe with a more negative DE pressure. As a consequence, CMB only posteriors hit the upper bound of the prior set at $H_0=\SI{100}{\km\per\s\per\mega\parsec}$, thus artificially shrinking cosmological constraints.} Adding LSS information becomes necessary and rewarding as it breaks the degeneracies of the constraints on $H_0$, $\ln 10^{10}A_s$, $\zre$ and $w$.
Constraints from the joint analysis ({${w = -1.04^{+0.21}_{-0.32}}$}) are consistent with a cosmological constant (${w = -1}$), while constraints from CMB favour a lower value of $w$.
We also note a strong correlation between the biases and $w$ of {79\%} and {92\%} for \textsf{CMASS} and \textsf{LOWZ} respectively.

In summary, in both cases, constraints from the joint analysis are substantially better for almost all parameters because of its ability to break degeneracies related to the chosen new parameters. This result constitutes a forceful encouragement to perform this type of analysis when data from the next generation of surveys becomes available.

\paragraph{Constraints on $w\mathrm{CDM}+\Smnu$.} 
\label{sub:constraints_on_w_mathrm_cdm_mnu_from_cmb_temperature_and_cmb_lensing_lss_correlations}

\begin{figure*}
\centering
	\includegraphics[width=17cm]{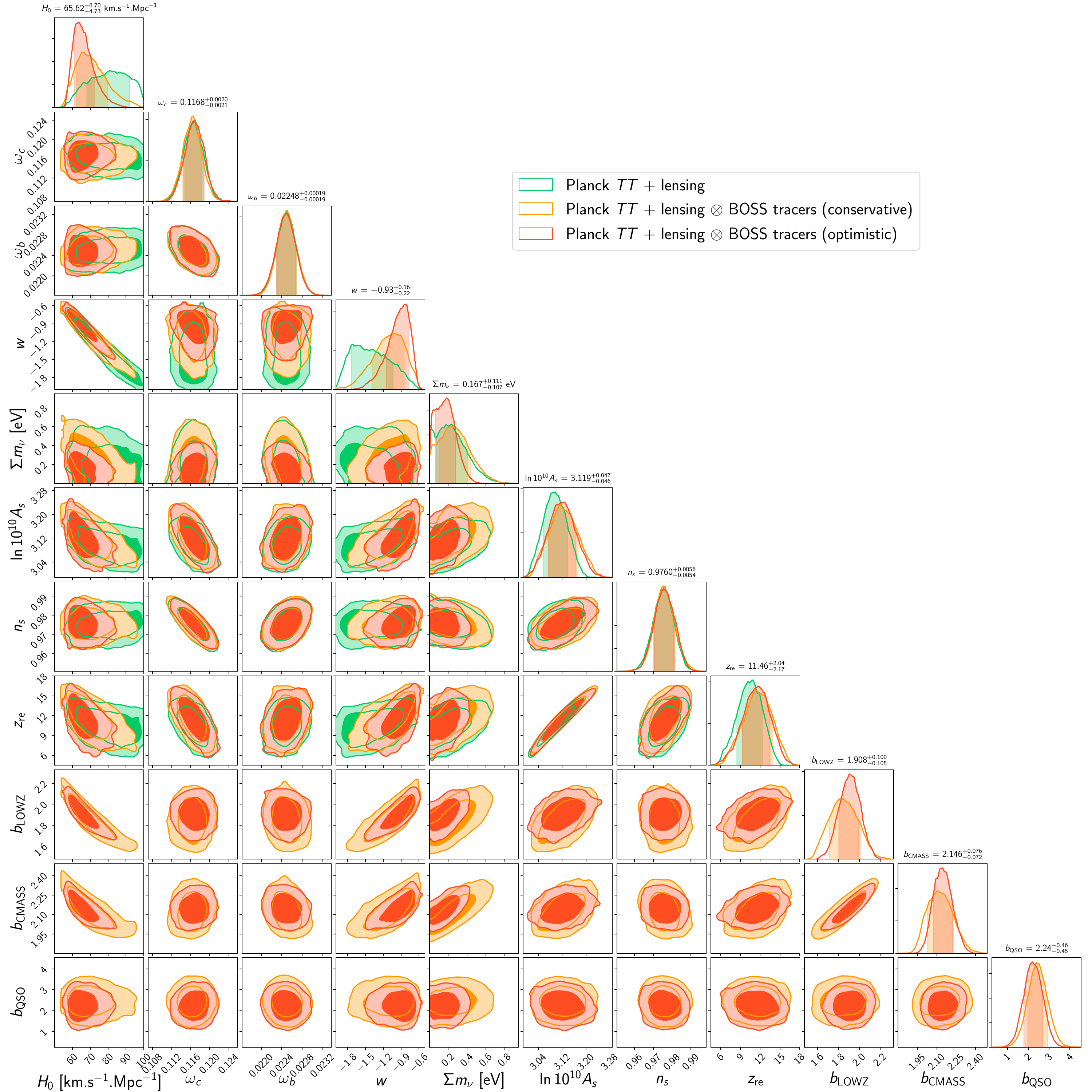}
	\caption{Constraints on the parameters of the $w\mathrm{CDM}+\Smnu$ model and tracer biases.}
	\label{fig:LCDM6params_w_mnu_corner}
\end{figure*}

\begin{figure}
	\includegraphics[width=\columnwidth]{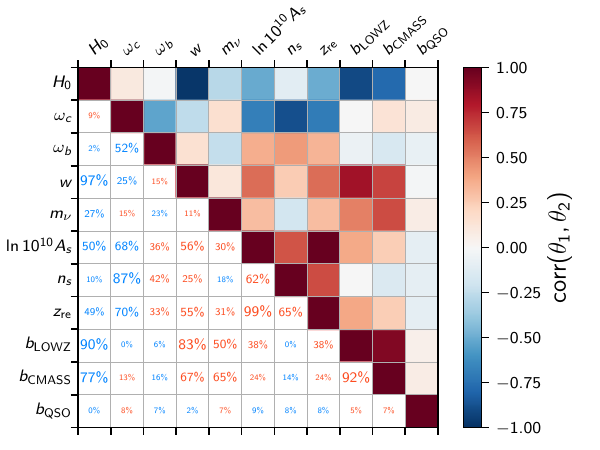}
	\caption{Correlation coefficient matrix of the $w\mathrm{CDM}+\Smnu$ model's parameters and biases from the joint analysis of CMB temperature and the correlations of CMB lensing and large-scale structure (see the constraints on \cref{fig:LCDM6params_w_mnu_corner}). The upper triangle is colour encoded, red (respectively blue) meaning complete correlation (anti-correlation) between parameters. The lower triangle is given in percentage, written in red (blue) for positive (negative) correlation.}
	\label{fig:corr_LCDMwm}
\end{figure}

\begin{figure}
	\includegraphics[width=\columnwidth]{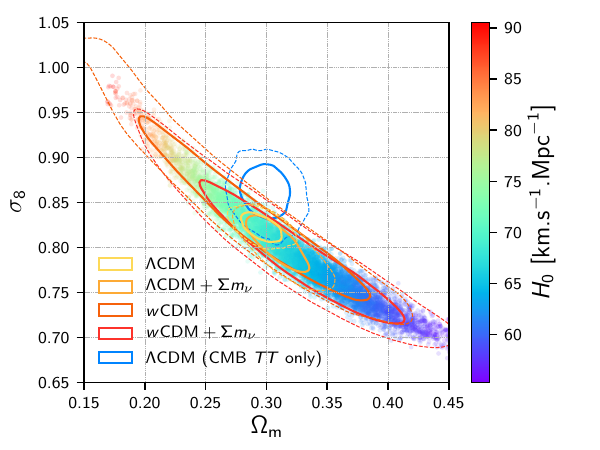}
	\caption{Confidence regions for $\sigma_8$ and $\Omegam$ from the joint analysis (optimistic cut) for the 6-parameter \LCDM model and its extensions to the total mass of neutrinos $\Smnu$ and the dark energy equation of state $w$ (from yellow to red). Constraints are compatible with each other across the tested models, though with increasing degeneracy. The coloured points are samples from the $w\mathrm{CDM}+\Smnu$ chain. The confidence region obtained from CMB $TT$ only for the \LCDM model is shown in blue for comparison.}
	\label{fig:xcorfull_PlanckTT_sigma8_Omegam}
\end{figure}

Finally, in the last set of MCMC analyses, we release both the total mass of neutrinos $\Smnu$ and the DE equation of state $w$ and demonstrate that a joint analysis of currently available data can constrain the 8-parameter cosmological model {$w$CDM + $\Smnu$}, with free parameters $H_0$, $\omegab$, $\omegac$, $\ln10^{10}\As$, $\ns$, $\zre$, $\Smnu$ and $w$. Similarly to the previous cases, the results are presented on \cref{fig:LCDM6params_w_mnu_corner} for the full joint analysis and for CMB data, allowing for comparison.
{As expected, CMB temperature data alone is found unable to constrain this model, and posterior distributions hit the prior upper bound on $H_0$. 
However, additional information extracted by the joint analysis (partially) breaks the $w - H_0$ degeneracy, enabling for control of all eight cosmological parameters, plus the biases, \emph{i.e.} eleven parameters in total. For these reasons and for readability of the figures, we only show the posterior distribution from ``Planck $TT$ + lensing'' data and the joint analysis.}

Both the conservative and optimistic cuts in the joint analysis yield constraints in agreement with the current picture of the \LCDM model as well as those obtained in the previous sections, with a value of {${w = -0.93^{+0.16}_{-0.22}}$} consistent with a cosmological constant. {The conservative cut does not improve the upper bound on the total neutrino mass, but it substantially improves constraints on $H_0$ and $w$ in comparison to those derived from CMB data alone.}
With the optimistic cut, we obtain a higher bound on the mass of the neutrinos of {${\Smnu < \SI{0.28}{\eV}\ [68\%]}$ (${< \SI{0.39}{\eV}\ [95\%]}$)} and a low value of the Hubble constant of {${H_0=\SI[parse-numbers=false]{65.6^{+6.7}_{-4.8}}{\km\per\s\per\mega\parsec}}$}, albeit with larger error bars with respect to the previous sections.
The correlation coefficient matrix reveals a strong correlation of galaxy biases with $H_0$, $w$ and $\Smnu$ (see \cref{fig:corr_LCDMwm}) indicating that upcoming surveys will require exquisite control of these biases to get tight constraints on $w$ and its possible time evolution.

Because of the degeneracy between $H_0$, $w$ and $\Smnu$, precision is lost on $\Omegam$, even though the physical density ${\omegam \equiv \Omegam h^2}$ is well constrained by CMB $TT$ and CMB lensing even in this model (we find $\omegam = 0.1411 \pm 0.0023$). In the ($\sigma_8,\;\Omegam$) plane (see \cref{fig:xcorfull_PlanckTT_sigma8_Omegam}), we obtain constraints that are consistent from the joint analysis over the models tested here, with increasing degeneracy. We measure {${\sigma_8^{2.7} \Omegam = 0.1709 \pm 0.0068}$} from the joint analysis on the {8-parameter} {$w$CDM + $\Smnu$} model.

\begin{table*}
    \centering
    \caption{Cosmological constraints from the full joint analysis of CMB temperature, CMB lensing and BOSS spectroscopic tracers with the optimistic cut (see \cref{sub:cross_correlation_data}). We test four different cosmological models and use the same flat priors in all cases. Constraints on the tracers' biases are also given. The last three rows give constraints on the derived parameters $\Omegam$, $\sigma_8$ and $\tau$. This table gives the median and asymmetric 68\% error bars.}
    \label{tab:constraints}
    \begin{tabular}{llllll}
        \hline
		Parameter [unit] & \LCDM & \LCDM + $\Sigma \mnu$ & $w$CDM & $w$CDM + $\Sigma \mnu$ & Prior (flat) \\
		\hline
		$H_0$  [\si{\km\per\s\per\mega\parsec}] & $68.77^{+0.87}_{-0.85}$ & $67.6^{+1.3}_{-1.4}$ & $69.7^{+10.2}_{-6.3}$ & $65.6^{+6.7}_{-4.8}$ & $[50,100]$ \\
		$\omega_{\rm b}$ & $\left( 225.1^{+1.8}_{-1.9} \right) \times 10^{-4}$ & $\left( 224.6\pm 1.8 \right) \times 10^{-4}$ & $\left( 224.8\pm 1.9 \right) \times 10^{-4}$ & $\left( 224.8\pm 1.9 \right) \times 10^{-4}$ & {$[0.01,0.03]$}\\
		$\omega_{\rm c}$ & $\left( 116.7\pm 1.9 \right) \times 10^{-3}$ & $\left( 117.2\pm 1.9 \right) \times 10^{-3}$ & $\left( 117.0\pm 2.2 \right) \times 10^{-3}$ & $\left( 116.8^{+2.0}_{-2.1} \right) \times 10^{-3}$ & {$[0.05,0.2]$}\\
		$\ln 10^{10} A_{\rm s}$ & $3.092^{+0.036}_{-0.034}$ & $3.104^{+0.038}_{-0.036}$ & $3.085^{+0.050}_{-0.049}$ & $3.119^{+0.047}_{-0.046}$ & $[2.5,3.5]$\\
		$n_{\rm s}$ & $\left( 976.2^{+5.0}_{-5.1} \right) \times 10^{-3}$ & $\left( 975.0^{+4.8}_{-5.2} \right) \times 10^{-3}$ & $\left( 975.3^{+5.8}_{-5.6} \right) \times 10^{-3}$ & $\left( 976.0^{+5.6}_{-5.5} \right) \times 10^{-3}$ & $[0.8,1.1]$\\
		$z_{\rm re}$ & $10.2^{+1.6}_{-1.7}$ & $10.8\pm 1.7$ & $9.9^{+2.3}_{-2.6}$ & $11.5^{+2.0}_{-2.2}$ & $[0,20]$ \\
		$\Sigma m_{\nu}$ [eV] & -- & $0.16^{+0.12}_{-0.10}$ & -- & $0.17\pm 0.11$ & $[0,10]$ \\
		$w$ & -- & -- & $-1.04^{+0.21}_{-0.32}$ & $-0.93^{+0.16}_{-0.22}$ & $[-2,0]$ \\
        \hline
		$b_{\lowz}$ & $1.837^{+0.034}_{-0.033}$ & $1.880^{+0.059}_{-0.051}$ & $1.826^{+0.100}_{-0.110}$ & $1.91^{+0.10}_{-0.11}$ & $[0,10]$ \\
		$b_{\cmass}$ & $2.086\pm 0.032$ & $2.130^{+0.058}_{-0.050}$ & $2.083^{+0.059}_{-0.055}$ & $2.146^{+0.077}_{-0.072}$ & $[0,10]$ \\
		$b_{\qso}$ & $2.19^{+0.45}_{-0.44}$ & $2.24^{+0.46}_{-0.45}$ & $2.21^{+0.44}_{-0.43}$ & $2.24\pm 0.46$ & $[0,10]$ \\
        \hline
		$\Omega_{\rm m}$ & $0.296\pm 0.011$ & $0.309^{+0.018}_{-0.015}$ & $0.288^{+0.058}_{-0.068}$ & $0.327^{+0.054}_{-0.058}$ & -- \\
		$\sigma_8$ & $0.822^{+0.011}_{-0.010}$ & $0.805^{+0.019}_{-0.023}$ & $0.830^{+0.082}_{-0.055}$ & $0.787^{+0.060}_{-0.046}$ & -- \\
		$\tau$ & $0.082^{+0.020}_{-0.019}$ & $0.088\pm 0.020$ & $0.078^{+0.028}_{-0.027}$ & $0.096^{+0.026}_{-0.025}$ & -- \\
		\hline
    \end{tabular}
\end{table*}


\subsection{Limits and perspectives} 
\label{sub:discussion}

In this section, we discuss assumptions that were made and technical difficulties that we were able to pinpoint.

We neglected the correlation, generated by the ISW effect, between the CMB temperature map and the large-scale structure as traced by CMB lensing or spectroscopic tracers. This correlation originates in the net energy gain (loss) of photons crossing gravitational potentials wells (hills) evolving thanks to dark energy. In principle, this would lead to underestimation of error bars on cosmological parameters. However, this correlation is weak and affects only very large scales ${\ell \lesssim 40}$, and it has not been detected with a strong statistical significance on SDSS galaxies: the signal-to-noise ratios for the correlation with the {\lowz} and {\cmass} samples reported by the Planck collaboration is of order 2.4, and that with the lensing map (corresponding to a temperature bispectrum) is of order 3.2 \citep{2016A&A...594A..21P}. Therefore, taking this cross-correlation term into account would not dramatically change our constraints.

One possible source of systematics in the galaxy-lensing cross-correlations comes from the tSZ component separation that is required to produce the lensing map \citep{2016A&A...594A...9P,2016A&A...594A..10P,2016A&A...594A..22P}. Free electrons in hot galaxy clusters imprint a specific local spectral distortion on the CMB temperature map. These clusters must be identified and removed before measuring the spatial distortion due to gravitational lensing. If these clusters hold some of the galaxies in the samples we use, this might lead to a systematic underestimation of the lensing signal in the direction of these galaxies. However, the SZ decrement from SDSS LRGs is small, as can be seen in Table~2 of \citet{2011ApJ...736...39H}. Moreover, the residual SZ signal primarily increases the noise in the lensing map and is unlikely to produce appreciable bias \citep[see the systematics checks in][]{2015PhRvL.114o1302M}.

\begin{figure}
	\includegraphics[width=\columnwidth]{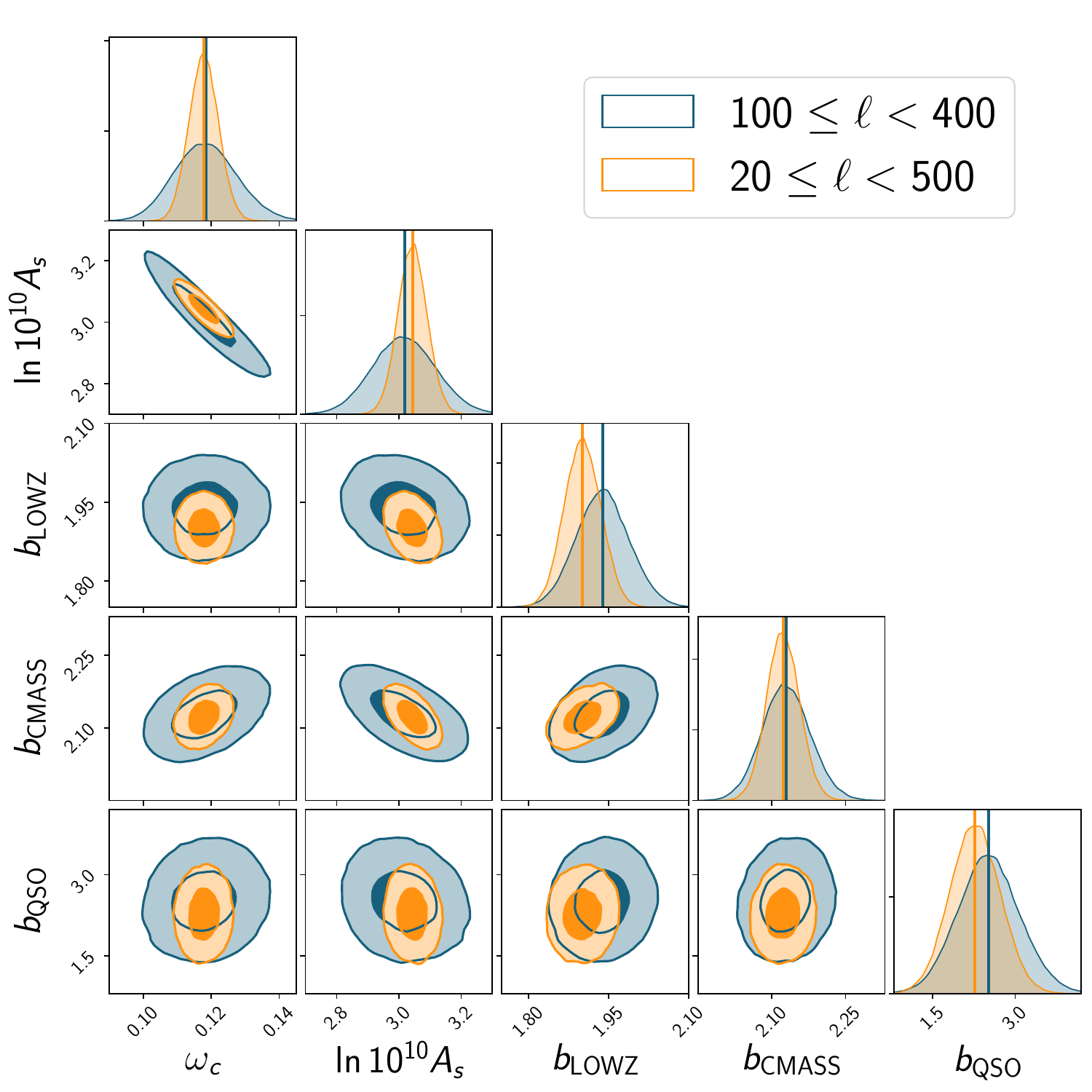}
	\caption{Comparison between the posterior distributions sampled by MCMC analysis using the CMB lensing-LSS correlations dataset for different multipole ranges: in orange, the constraints resulting from the full multipole range ($20 \leq \ell < 500 $) used in the rest of this work, and in blue those resulting from a smaller multipole range excluding small and large scales ($100 \leq \ell < 400 $). The distribution are compatible, which means that if there is a modelling issue either at large scales (\emph{e.g.} due to the RSD) or at small scales (\emph{e.g.} due to uncertainties in the non-linear matter power spectrum), it does not dramatically affect the constraints on cosmological parameters.}
	\label{fig:xcorfull_LCDM2params_test_ellrange_corner}
\end{figure}


We now discuss theoretical uncertainties. The first one comes from the Limber approximation \citep{2008PhRvD..78l3506L}: it fails at very large scales, the transition scale depending on the width of the redshift bin considered \citep{2017A&A...602A..72C}. In this work, we used very broad redshift bins and discarded low multipoles $\ell < 20$, so as to be in the safe regime of the approximation. We also did not consider redshift space distortion (RSD) for the same reasons{, as they were shown to be negligible for $\ell\gtrsim20$ \citep{2015ApJ...814..145A,Padmanabhan:2007ci} and rapidly decreasing with the width of redshift bins \citep{Saito:PgNvTFaY}}. To test that RSD or other large-scale effects were not driving parameter constraints inconsistently, we performed an MCMC analysis of CMB lensing-galaxy correlations on a smaller multipole range (${100 \leq \ell < 400}$, see \cref{fig:xcorfull_LCDM2params_test_ellrange_corner}) and found no significant deviation. However, future surveys aiming at measuring extremely large-scales \citep{2015PhRvD..92f3525A} will require better modelling, especially for tomographic studies with thin redshift bins.

In this paper, we used a Gaussian likelihood and a Gaussian covariance, \emph{i.e.} we did not incorporate higher order statistics of the matter density field nor the so-called super-sample variance due to the finite size of the surveyed volume and inaccessible modes therein \citep{2014PhRvD..90l3523S,2017MNRAS.470.2100K,2017PhRvD..95l3512S}. At the current level of signal-to-noise ratio, these simplifications are safe\footnote{{Super-sample variance, for instance, can introduce correlations between multipoles of order 10\%  at $z=0.1$ and decreasing with redshift and multipole number, see \citet{2016JCAP...08..005L}.}} but they should be lifted in future data analysis. One limitation of our method regarding the covariance matrix is that the computation of the $\mathbfss{X}$ and $\mathbfss{Y}$ matrices, even if it needs to be done only once, is numerically expensive since they grow linearly with the multipole range but as $n^4/4$ with the number $n$ of different masks, and the estimate remains noisy far from the diagonal. However, this method has the advantage of naturally taking care of partial sky coverage, without the need of inverting the mixing matrix to recover full-sky spectra, necessarily introducing numerical noise in the data.
Devising a method that takes care of partial sky coverage whilst incorporating all relevant non-Gaussian terms will be an important task for future surveys \citep{2016arXiv161205958L}.

Finally, the non-linear power spectrum of the matter density field suffers from theoretical uncertainties \citep{Baldauf:2016tk}. {Throughout this analysis, we have used a version of the \texttt{halofit} model that includes the effect of massive neutrinos and predicts the matter power spectrum past the non-linear transition scale (around $k_{\mathrm{nl}} \sim \SI{0.1}{\h\per\mega\parsec}$). So as to be agnostic, we considered two cuts at small scales at \SI{0.1}{\per\mega\parsec} and \SI{0.15}{\per\mega\parsec}. The \texttt{halofit} model reaches percent agreement with simulations at these scales and we have demonstrated that we recover consistent constraints for both cuts, with small shifts of best-fit values for all parameters, though expectedly tighter in the optimistic case.
The scientific gain from near future surveys in terms of cosmological constraints will depend on our ability to model these non-linearities.} In particular, the suppression of power due to massive neutrinos \citep{2006PhR...429..307L} and the contribution of baryonic and feedback processes \citep{2017MNRAS.467.3024L} at these scales will certainly be an important theoretical issue for future surveys.


\section{Conclusion} 
\label{sec:conclusion}

Cosmological experiments carried out in the last few decades have enabled the construction of the \LCDM model. In this picture, cold dark matter drives the formation of the large-scale structure of the Universe and dark energy fuels the recent accelerated expansion. The combination of independent observations, such as the map of the anisotropies of the cosmic microwave background, distances of type IA {supernov\ae} and the measurement of the scale of the baryon acoustic oscillations, have set constraints on the content of the Universe.
However, the analysis of currently available data cannot distinguish between various models of dark matter and dark energy. Going further and deciphering the nature of these components requires better constraints, and thus, more information. To this end, deep galaxy surveys -- such as LSST, Euclid and WFIRST -- and CMB imagers -- such as CMB-S4 and the Simons Observatory -- with wide sky coverage and high resolution are currently under development. In the coming decade, they will probe the matter density field with ground-breaking precision and significantly increase the amount of cosmological information. Independent cosmological analyses have a strong potential to reveal new science, but model comparison will rely on exhausting the cosmological information held in the measurements of different cosmic probes and all their cross-correlations. In other words, {joint analyses of these probes are required, thus explaining the recent intense activity around this subject (see the introduction)}.

In this paper, we have presented a joint analysis of currently available data combining CMB measurements --both temperature anisotropies and gravitational lensing-- from the Planck satellite and LSS tracers from the SDSS-III/BOSS spectroscopic survey{, taking advantage of the large areas covered by these surveys and their large overlap (a requirement for measuring cross-correlations with a high signal-to-noise ratio)}. To this end, we developed a general framework in \texttt{NumCosmo} to compute and analyse the auto and cross-correlations between an arbitrary number of cosmological probes, which is publicly available. In particular, we applied our framework to analyse CMB lensing and galaxy clustering at once by measuring all relevant auto and cross angular power spectra (shown in \cref{fig:clobs}).
{We validated the likelihood and pseudo-spectra estimators in \cref{sec:validation,sub:null_tests}. Note that our approach required few simplifications on the CMB lensing auto power spectrum (as explained in \cref{sub:Planck_data}), but we demonstrate in \cref{sec:lensing_auto_power_spectrum} that our pipeline yields unchanged cosmological constraints, except for a small, yet non-negligible, $0.5\sigma$ shift in the $\As-\zre$ degeneracy, below the level of statistical errors nonetheless.}
In \cref{ssub:constraints_from_cmb_lensing_large_scale_structure_correlations}, we showed how including cross-correlation information -- already present in the data -- improves constraints on cosmological parameters and decreases the statistical errors, for example, by 10\% for ${\ln 10^{10} \As}$ and 20\% for $\omegac$ (when other parameters are fixed). This highlights the fact that ignoring part of the cosmological information (in this case, the cross-correlations) could lead to inaccurate posterior distributions of the parameters.

{Next, we included CMB temperature {anisotropies information by adding the likelihood of the $C_{\ell}^{TT}$ power spectrum (thus neglecting the small ISW-induced CMB-LSS correlation as discussed in \cref{sub:discussion})} and carried out different MCMC analyses to constrain the base, {6-parameter}, flat \LCDM model. Finally, we explored constraints on the total mass of neutrinos and the DE equation of state, constraining four different cosmological models (see \cref{fig:LCDM6params_corner,fig:LCDM6params_extensions,fig:LCDM6params_w_mnu_corner}). We compared the performance of the joint analysis (using two different cuts) with analyses using only CMB data.
As expected, constraints from the joint analysis are stronger than those obtained from CMB data only, in all cases.
Because of the sensitivity of galaxy clustering and the CMB lensing-galaxy cross-correlations to $\Smnu$ and $w$, we were able to study extended models and constrain up to eight cosmological parameters at once (that is, $H_0$, $\omegab$, $\omegac$, $\As$, $\ns$, $\zre$, $\Smnu$ and $w$), which is impossible with either of the data sets considered separately.
}

As a result, we observe the (partial) breaking of several degeneracies and significantly better constraints for various parameters, although this depends upon exactly which parameters are constrained and which are assumed to be fixed.
We thus obtained upper limits on the total mass of neutrinos of \SI{0.28}{\eV}~[68\%] as a result of its impact on galaxy clustering at small scales, which is similar to limits obtained with other comparable analyses. {It is interesting to note here that combining CMB and BAO distance measurements currently yields tighter constraints on the total neutrino mass because BAO reconstruction includes more $k$-modes than power spectrum measurements that exclude non-linear scales \citep{2017PhRvD..96l3503V,2016MNRAS.457.1770C}. Better modelling of the power spectrum and clustering biases at these scales is thus a key to improve cosmological constraints. In addition, galaxy clustering information within the joint analysis enabled us to obtain constraints on the dark energy equation of state $w$, found to be consistent with $w=-1$.
We have also identified strong correlations between clustering biases and some cosmological parameters, in particular $H_0$ and $w$.
A downside is that future surveys will have to measure and marginalise over these biases with great precision in order to pin down the values of these parameters and to constrain a possible time dependence of dark energy. Interestingly, if we used a value of ${H_0=\SI{72}{\km\per\s\per\mega\parsec}}$ consistent with distance measurements from type Ia supernov\ae, then our constraints would favour a value of the DE equation of state of ${w\sim-1.1}$, \emph{i.e.} a phantom dark energy, which is disfavoured by theoretical considerations.
}

The strength of the cosmological constraints derived here is naturally limited by theoretical uncertainties in the non-linear regime on the matter power spectrum and linear clustering biases, as discussed in \cref{sub:discussion}. {To address this issue,} we have tested two different cut-off scales and shown that it results in consistent cosmological constraints across all models tested, although small shifts are expectedly observed. For instance, the higher bound on the total mass of neutrinos decreases when including smaller scales that are more strongly impacted by neutrinos. {Additionally, we have discussed in \cref{sub:discussion} several other limitations, such as the impact of the ISW effect, potential contamination of cross power spectra with CMB lensing by the SZ effect and the impact of the Limber approximation and redshift-space distorsions.}

In this work, {we followed an approach similar to \citet{2016PhRvD..94h3517N,2017PhRvD..95h3523N} that, however, used photometric data from the SDSS and DES, combined with geometric probes. Our approach is based on pseudo spectra with different masks for each observables and therefore exploits the full observed area for each probe, thus maximizing the signal-to-noise ratio of power spectra. We used a semi-analytic, cosmology-dependent covariance matrix that is less noisy than Monte Carlo estimates, but required the assumption that the fields we measure are Gaussian distributed. A unified framework that incorporates non-Gaussian terms in the covariance and handles partial sky coverage remains to be derived.}
{
We focused on spectroscopic observations of galaxies and quasars, insulating us from uncertainties inherent to photometric redshifts. However, we did not use galaxy weak lensing measurements (as was done in \citet{2016PhRvD..94h3517N}), a powerful probe of dark energy that will be measured by future deep surveys like LSST, Euclid and WFIRST. The trade-off between the precision of photometric redshifts and the much larger number of galaxies combined to this additional probe will certainly lead to even better results. Combining CMB lensing, galaxy lensing and galaxy clustering (both photometric and spectroscopic) in a fully joint analysis is a promising avenue for cosmological parameters estimation.}

Finally, in this near-future scenario of large amounts of data and joint analyses, we will be able to study different cosmological models emerging from different theories of gravity, such as effective field theories of dark energy \citep{2016JCAP...02..056G} or non-local gravity \citep{2016JCAP...05..068D}, and hopefully start to distinguish and rule out some models with strong statistical significance.

\section*{Acknowledgements}

CD would like to thank David Spergel and Emmanuel Schaan for useful comments and the opportunity to present a preliminary version of this work, Dhiraj Hazra for useful comments on the manuscript, Antony Lewis for useful comments about the lensing power spectrum, Alexie Leauthaud, Martin Kunz, Fabien Lacasa, Michele Maggiore, Mathew Madhavacheril and Sunny Vagnozzi for interesting discussions and comments.

The authors thank the anonymous referee for their thorough review, which resulted in major improvements, making our conclusions and cosmological constraints more robust.

The authors acknowledge use of \texttt{Healpy} \citep{2005ApJ...622..759G}, \texttt{CLASS} \citep{2014JCAP...09..032L} and the Planck likelihood codes \texttt{Plik}, \texttt{Commander} and \texttt{LensLike} available on the Planck Legacy Archive (\url{http://pla.esac.esa.int/pla/}).

MPL received support from the National Council for Scientific and Technological Development - Brazil (CNPq grant 202131/2014-9 and PCI/MCTIC/CBPF program), and from the Labex ENIGMASS.

SDPV acknowledges the financial support from CNPq (PCI/MCTIC/CBPF program) and BELSPO non-EU postdoctoral fellowship.

Funding for SDSS-III has been provided by the Alfred P. Sloan Foundation, the Participating Institutions, the National Science Foundation, and the U.S. Department of Energy Office of Science. The SDSS-III web site is \url{http://www.sdss3.org/}.

SDSS-III is managed by the Astrophysical Research Consortium for the Participating Institutions of the SDSS-III Collaboration including the University of Arizona, the Brazilian Participation Group, Brookhaven National Laboratory, Carnegie Mellon University, University of Florida, the French Participation Group, the German Participation Group, Harvard University, the Instituto de Astrofisica de Canarias, the Michigan State/Notre Dame/JINA Participation Group, Johns Hopkins University, Lawrence Berkeley National Laboratory, Max Planck Institute for Astrophysics, Max Planck Institute for Extraterrestrial Physics, New Mexico State University, New York University, Ohio State University, Pennsylvania State University, University of Portsmouth, Princeton University, the Spanish Participation Group, University of Tokyo, University of Utah, Vanderbilt University, University of Virginia, University of Washington, and Yale University.

Based on observations obtained with Planck (\url{http://www.esa.int/Planck}), an ESA science mission with instruments and contributions directly funded by ESA Member States, NASA, and Canada.




\bibliographystyle{mnras}
\bibliography{/Users/doux/Cosmo/biblio/papers,refsDoux}




\appendix

\section{Lensing auto-power spectrum} 
\label{sec:lensing_auto_power_spectrum}

\begin{figure}
    \centering
    \includegraphics[width=\columnwidth]{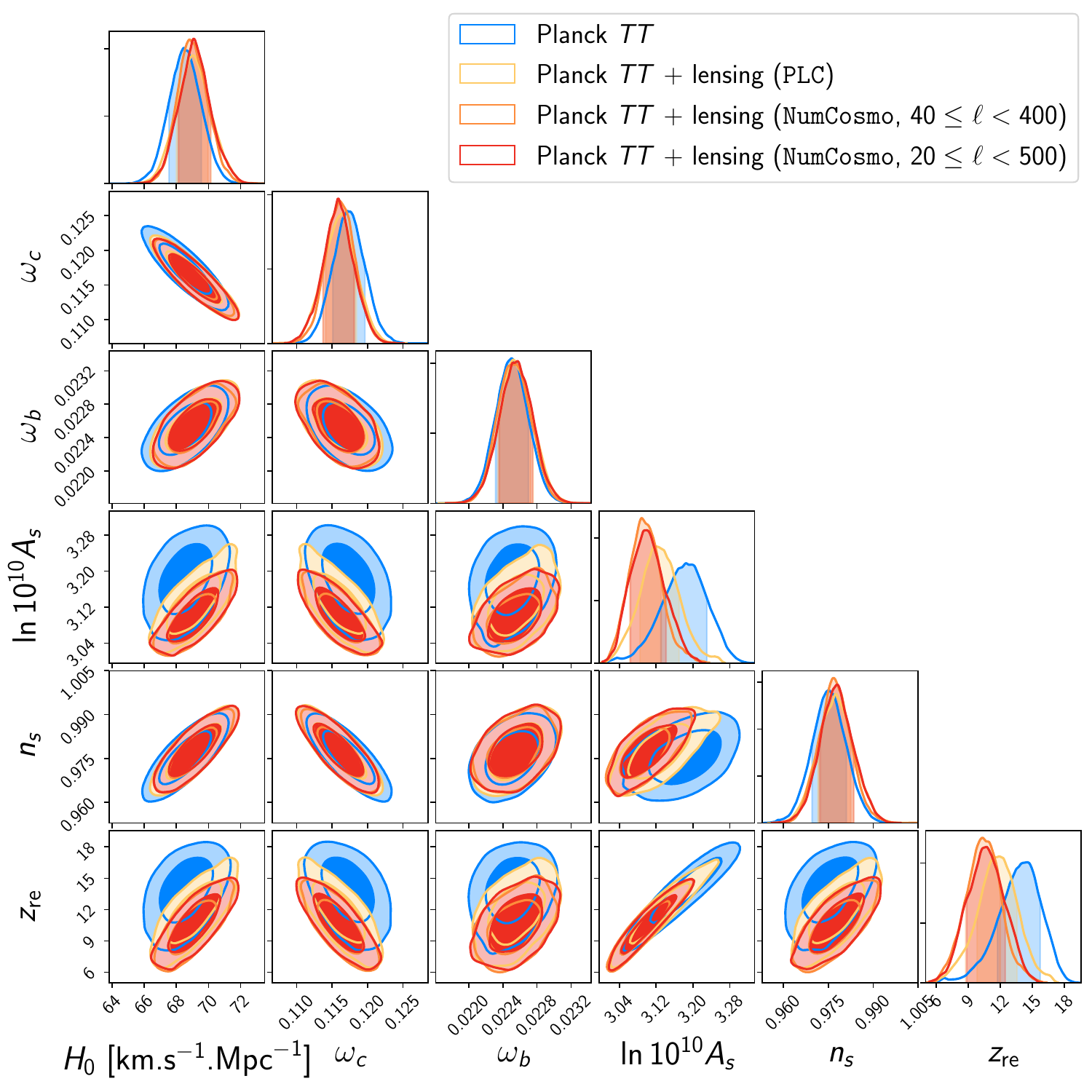}
    \caption{Comparison of the constraints derived with the Planck lensing likelihood code and data with our approach and likelihood based on \texttt{NumCosmo}.}
    \label{fig:lensing_compare}
\end{figure}

{
Throughout this work, we have used the CMB lensing auto-power spectrum from Planck based on the released reconstructed convergence map, neglecting corrections that were applied in the Planck cosmological analysis but that are not yet available in a user-friendly format. Our method is based on pseudo spectra while the Planck collaboration used full-sky spectra, making it difficult to quantify the impact of each correction.
Therefore, we compare cosmological constraints derived from CMB temperature and lensing, using either the Planck lensing likelihood code or our code based on \texttt{NumCosmo} using the conservative cut, $40\leq\ell<400$, and the optimistic cut, $20\leq\ell<500$, used throughout the rest of the analysis. We use the ensemble MCMC sampler of \texttt{NumCosmo} and show constraints on parameters of the base \LCDM model in \cref{fig:lensing_compare}. We find compatible constraints for the two codes, with very close maximum likelihood values, statistical uncertainties and shifts with respect to CMB $TT$ alone, except for a $0.5\,\sigma$ shift along the $\As e^{-2\tau}$ degeneracy with an impact on $\As$ and $\zre$ (and thus $\sigma_8$ and $\tau$). This demonstrates that our likelihood code performs very well, but that reconstruction corrections applied to the estimated lensing power spectrum have a non-negligible effect, that is however smaller than statistical errors. We conclude that our constraints on $\As$ and $\zre$ might be slightly biased, though at a reasonable level that does not impact our main conclusions.
}


\section{$\mathbfss{X}$/$\mathbfss{Y}$ matrices in the covariance} 
\label{sec:XY_matrices_in_the_covariance}

\begin{figure*}
\centering
	\includegraphics[width=17cm]{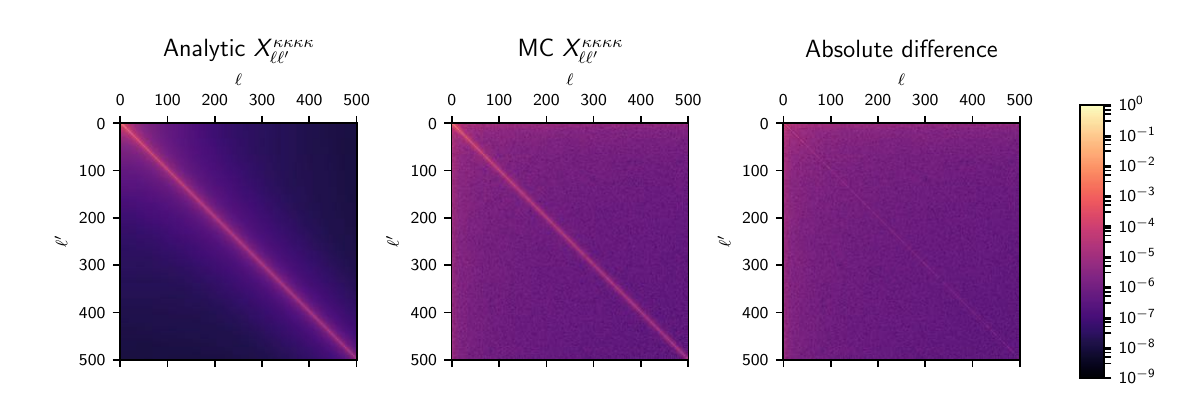}
	\caption{Comparison between the analytically-estimated (left panel) and simulation-estimated (middle panel) ${\mathbfss{X}_{\ell \ell'}^{\kappa \kappa \kappa \kappa}}$ matrices in logarithmic scale. The absolute difference is shown on the right panel. The important features are well captured: the precision is better than 2\% on the diagonal and degrades when getting further away from the diagonal. The middle panel shows that the far off-diagonal terms are dominated by numerical noise from our MC simulations, but are four orders of magnitudes smaller than the diagonal terms which are the most important, making it safe to use in the covariance matrix. The right panel shows the absolute difference.}
	\label{fig:XY}
\end{figure*}

The $\mathbfss{X}$ and $\mathbfss{Y}$ matrices appearing in eq.~\eqref{eq:covariance} have the following analytical expressions \citep{2005MNRAS.360.1262B},
\begin{equation}
    \begin{split}
    	&\mathbfss{X}_{\ell_1 \ell_2}^{ABCD} = \frac{1}{(2\ell_1+1)(2\ell_2+1)} \times \\ & \sum\limits_{m_1 m_2} \sum\limits_{\ell_3 m_3 } \sum\limits_{\ell_4 m_4} W^{\ A}_{\ell_1 \ell_3 m_1 m_3} \overline{W}^{\ B}_{\ell_2 \ell_3 m_2 m_3} W^{\ C}_{\ell_2 \ell_4 m_2 m_4} \overline{W}^{\ D}_{\ell_1 \ell_4 m_1 m_4}&\\
    	&\mathbfss{Y}_{\ell_1 \ell_2}^{ABCD} = \frac{1}{(2\ell_1+1)(2\ell_2+1)} \times \\ & \sum\limits_{m_1 m_2} \sum\limits_{\ell_3 m_3 }  \sum\limits_{\ell_4 m_4} W^{\ A}_{\ell_1 \ell_3 m_1 m_3} \overline{W}^{\ C}_{\ell_2 \ell_3 m_2 m_3} W^{\ B}_{\ell_2 \ell_4 m_2 m_4} \overline{W}^{\ D}_{\ell_1 \ell_4 m_1 m_4},
    \end{split}
\end{equation}
where the $W^{\ A}_{\ell \ell' m m'}$ describe the convolution of the mask ($\overline{W}^{\ A}_{\ell \ell' m m'}$ is its complex conjugate), \emph{i.e.} if the field $A(\hatn)$ has full-sky spherical harmonics coefficients $A_{\ell m}$ and pseudo-coefficients $\tilde{A}_{\ell m}$ then
\begin{equation}
	\tilde{A}_{\ell m} = \sum_{\ell' m'} W^{\ A}_{\ell \ell' m m'} A_{\ell m}.
\end{equation}

These cannot be analytically computed and MC simulations are therefore required. We take advantage of the fact that eq.~\eqref{eq:covariance} is exact if initial full-sky spectra do not depend on $\ell$ and that they need not have physically relevant values. The algorithm then proceeds as follows. First, we generate sets of four correlated maps with generic constant input auto and cross spectra, which we mask by the four masks used in our analysis. We then compute the spectra of the masked maps, thus building a collection of estimated pseudo spectra $\{\tilde{C}_{\ell}^{AB, i} \}_i$ where $i$ represents the simulation index. The empirical covariance of the set of pseudo spectra is finally computed.
Knowing the input spectra, an estimate of $\mathbfss{X}_{\ell \ell'}^{ABCD}$ and $\mathbfss{Y}_{\ell \ell'}^{ABCD}$ can be obtained using eq.~\eqref{eq:covariance}. In the case where ${A = B}$ or ${C = D}$ and only in this case, the terms in the square roots in eq.~\eqref{eq:covariance} are equal and $\mathbfss{X}_{\ell \ell'}^{ABCD}$ and $\mathbfss{Y}_{\ell \ell'}^{ABCD}$ cannot be distinguished, but for all the other cases, it requires two sets of simulations to disentangle them.

We estimate the error on the empirical covariance matrices by bootstrapping the pseudo spectra $\{\tilde{C}_{\ell}^{AB, i} \}_i$ and require that the ratio of the norms of the error matrix to that of the empirical covariance matrix is smaller than 1\%, which in our analysis necessitated more than 200~000 simulations.

In the case where ${A=B=C=D}$, these matrices reduce to symmetrized mixing matrices
\begin{equation}
	\mathbfss{X}_{\ell \ell'}^{AAAA} = \mathbfss{Y}_{\ell \ell'}^{AAAA} = \frac{1}{2\ell'+1}\mathbfss{M}_{\ell \ell'}^{AA},
\end{equation}
which allows for comparison and validation of the MC simulations (see \cref{fig:XY}). We find percent-level agreement on the diagonal, with a decreasing precision when moving further away from the diagonal as numerical noise (at least four orders of magnitude smaller than the diagonal elements) starts dominating.

\section{Profile likelihood} 
\label{sec:profile_likelihood}

\begin{figure}
	\includegraphics[width=\columnwidth]{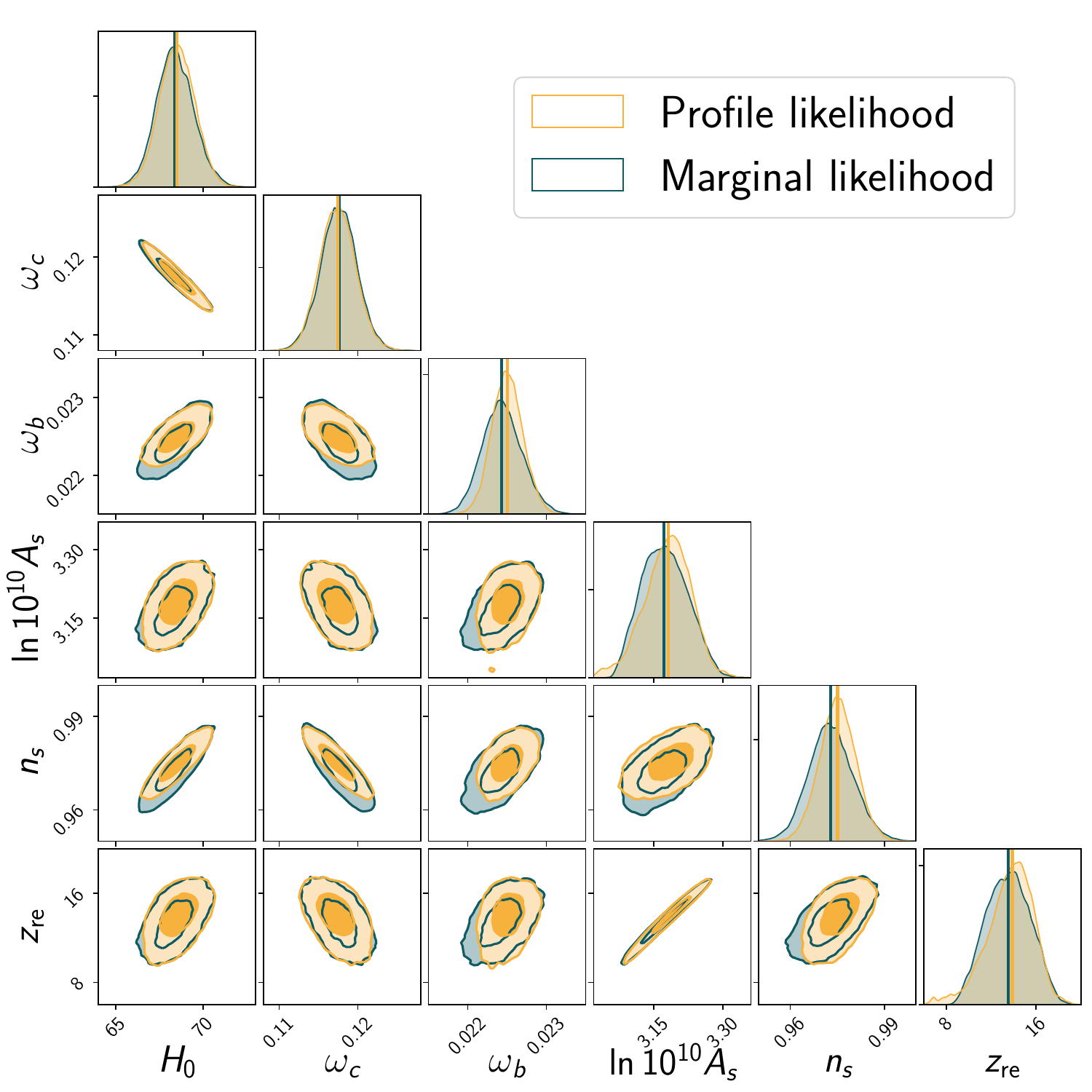}
	\caption{Comparison between the marginal likelihood (dark blue) and the profile likelihood (yellow). On the diagonal, the one-dimensional projections of the posterior distribution for each parameter is shown, the vertical lines corresponding to the mean value. For all the parameters, the difference between the means of the two distribution is much smaller than the statistical error. The standard deviations are also very close, with at worst a 10\% decrease for for $\ns$ and $\omegab$.}
	\label{fig:PlanckTT_LCDM6params_testsubfit_corner}
\end{figure}

In order to accelerate our MCMC analyses, we choose to use the profile likelihood instead of the marginal likelihood for the nuisance parameters. The reason is that this procedure decreases the dimension of the parameter space and requires less calls to the Boltzmann code, resulting in an overall faster convergence of the posterior distribution of the cosmological parameters. In practice, it amounts to compute the maximum likelihood estimator value of the nuisance parameters $\hat{\bmath{A}} (\bmath{\theta})$ for each set of cosmological parameters $\bmath{\theta}$ given the data (which is fast), and use this value in the likelihood. The posterior distribution is then given by
$\mathcal{L}_{\mathrm{profile}}\left( \bmath{\theta} \middle\vert C_{\ell}^{TT} \right) \propto \mathcal{L}\left( C_{\ell}^{TT} \middle\vert \bmath{\theta}, \hat{\bmath{A}} (\bmath{\theta}) \right)$
while the marginal likelihood is
$\mathcal{L}_{\mathrm{marginal}}\left( \bmath{\theta} \middle\vert C_{\ell}^{TT} \right) \propto \int \mathcal{L}\left( C_{\ell}^{TT} \middle\vert \bmath{\theta}, \bmath{A} \right) \mathrm{d}\bmath{A}$.
We demonstrate that it doesn't affect the results on the cosmological parameters by running two MCMC using only the CMB temperature power spectrum $C_{\ell}^{TT}$, one performing the nuisance parameters subfitting procedure and the other using the standard marginalisation procedure. \cref{fig:PlanckTT_LCDM6params_testsubfit_corner} shows the posterior distribution in these two cases. The mean value of each parameter in both runs is shown in the one-dimensional plots on the diagonal. In all cases, the variation of the mean is much smaller than the statistical variance, and the standard deviation is at worst decreased by 15\% in the profile likelihood (for $\ns$ and $\omegab$, two parameters which are poorly constrained by the other observations), with almost no difference for the other parameters. This indicates that we can use either likelihoods indifferently. Since the profile likelihood method is faster overall, and that we don't have other nuisance parameters, we used it for all simulations in \cref{ssub:cosmological_constraints_from_the_joint_analysis_of_planck_and_boss_data}.


\section{The \texttt{NumCosmo} library}
\label{sec:ncxcor}

In this section, we provide a short description of the Numerical Cosmology library (\texttt{NumCosmo}, available on GitHub\footnote{\url{https://numcosmo.github.io/}}, \citet{2014ascl.soft08013D}). Apart from the crude data (observational maps), for which we used some \texttt{Healpix} functions to generate the observed pseudo-$C_\ell^{AB}$ values, as mentioned in \cref{sec:Data}, all other pieces of the pipeline made use of \texttt{NumCosmo}. For a complete description we refer the reader to \citet{Vitenti2018}.

\texttt{NumCosmo} contains a comprehensive set of tools to compute cosmological observables and to perform statistical analysis. The library is written in C, but since it uses the \texttt{GObject} framework\footnote{\url{https://developer.gnome.org/gobject/stable/}}, it is developed in a object-oriented fashion. Additionally, it has automatic bindings for every language that supports \texttt{GObject} introspection (\emph{e.g.} Python, Ruby or Perl).

Physical models are implemented via the abstract class \texttt{NcmModel}. In particular, the \LCDM and $w$CDM models, and all respective relevant functions are implemented in \texttt{NcHICosmoDE} and child classes (such as \texttt{NcHICosmoDEXcdm}), the primordial power spectrum is implemented in \texttt{NcHIPrim}, the reionization model in \texttt{NcHIReion}. Data objects deriving from the abstract class \texttt{NcmData} encapsulate the observations and implement likelihood functions. A general object for statistical analysis \texttt{NcmFit} is then built from the data and the model.

We first address the computation of the theoretical angular power spectrum, $C_\ell^{AB}$ (see eqs.~\eqref{eq:cl_1} and \eqref{eq:cl_general}) and the likelihood function eq.~\eqref{eq:likelihood}:
\begin{enumerate}
    \item \texttt{NcXCor}: abstract class that comprises, among others, the methods to compute the auto and cross power spectra $C_\ell^{AA}$ and $C_\ell^{AB}$.
    \item \texttt{NcXCorLimberKernel}: abstract class of the type \texttt{NcmModel}\footnote{Being a \texttt{NcmModel}, each implementation of \texttt{NcXCorLimberKernel} can define a respective set of parameters. For instance, the linear bias, $b(z)$, in eq.~\eqref{eq:Wg}.} which defines the methods and general properties that a kernel $W^A(z)$ must implement, for any observable $A$. For instance, the computation of $W^A(z)$ at a given $z$ and for a set of cosmological parameters, and the number of multipoles to be calculated.
    \begin{itemize}
        \item \texttt{NcXCorLimberKernelCMBLensing}: implements the CMB lensing kernel, $W^\kcmb(z)$ [eq.~\eqref{eq:Wk}].
        \item \texttt{NcXCorLimberKernelGal}: implements the galaxy kernel, $W^g(z)$ [eq.~\eqref{eq:Wg}].
    \end{itemize}
    \item \texttt{NcPowSpecMNL}: abstract class for the nonlinear matter power spectrum. Here we use the \texttt{halofit} approach \citep{2003MNRAS.341.1311S,2012ApJ...761..152T}, which we implemented in the \texttt{NcPowspecMNLHaloFit} class. The linear matter power spectrum is calculated using \texttt{NcPowspecMLCBE}, i.e., the \texttt{NumCosmo} backend for the Cosmic Linear Anisotropy Solving System (CLASS) \citep{2011JCAP...09..032L}.
    \item \texttt{NcDataXCor}: this object builds the likelihood given by eq.~\eqref{eq:likelihood}, and it derives from \texttt{NcmDataGaussCov}, \emph{i.e.}, the object that describes Gaussian-distributed data with non-diagonal covariance matrix.
\end{enumerate}

Regarding the statistical analyses performed in this work, we made use of the following \texttt{NumCosmo} tools:
\begin{enumerate}
\item \texttt{NcmFit}: implements various interfaces with best-fit finders. In this work, we used the interface with the NLOpt library\footnote{\url{http://ab-initio.mit.edu/nlopt}} and the Nelder-Mead algorithm.
\item \texttt{NcmFitMC}: implements the Monte Carlo analysis described in \cref{sec:validation}, using the same best-fit finder.
\item \texttt{NcmFitESMCMC}: implements the Ensemble Sampler Markov Chain Monte Carlo \citep{Goodman:2010et} analysis used throughout \cref{sec:analyses_and_results} of this paper. It requires an initial sampler \texttt{NcmMSetTransKern} and another sampler to move the walkers \texttt{NcmFitESMCMCWalker}.
\end{enumerate}

Finally, we used CMB temperature data from Planck:
\begin{enumerate}
    \item \texttt{NcPlanckFICorTT}: implements Planck foreground and instrumental models for TT measurements.
    \item \texttt{NcDataPlanckLKL}: implements the interface with Planck's likelihood codes \texttt{Plik} and \texttt{Commander}.
\end{enumerate}

\section{MCMC convergence tests} 
\label{sec:mcmc_convergence_tests}

In this work, we checked the convergence of the MCMC chains using three different methods, which we implemented closely following the \texttt{R} package \texttt{CODA}~\citep{Plummer:2006uo}. The first is the Multivariate Potential
Scale Reduction Factor~\citep[MPSRF,][]{1992StaSc...7..457G, Brooks:2012ju}. This
method requires multiple chains, whose initial values must be
over-dispersed in comparison with the posterior, and quantifies the mixing of the walkers by comparing the ensemble variance to the per-walker variance.
Nevertheless, we do not know \emph{a priori} the posterior and, for this
reason, we may only guess what an over-dispersed distribution would be.

The second method is the Heidelberger-Welch diagnostic test
\citep{Heidelberger:1981ih, Heidelberger:1983ia}, which consists in applying
the Schruben stationarity test~\citep{Schruben:1982iq} to subsets of a chain to obtain one that satisfies the test for a given $p$-value. Since we are using an
ensemble sampler, we can apply this test to each individual chain, or,
more efficiently, to the ensemble mean of each parameter. We applied the
individual approach only when the Markov chain presents convergence
problems.

In the third approach, we calculated the autocorrelation time as
proposed by \citet{Goodman:2010et}. However, in \texttt{NumCosmo},
instead of estimating the autocorrelation time directly from the
autocorrelations, we fit an Auto Regressive (AR) model as in
\texttt{CODA}. In the AR model fitting, we use the bias corrected Akaike
Information Criterion (AICc) \citep{HURVICH:1989ev} to choose the best AR
order to use for a given parameter in a chosen chain. This provides a
less noisy estimate of the autocorrelation time than the direct inference
from the autocorrelations \citep[see][]{Goodman:2010et}. The Effective Sample Size (ESS) is computed using that estimated
autocorrelation time and
provides an equivalent measure of the effective number of independent points in each chain. Finally, the variance of the sample mean of the parameters is given by the empirical variance of the sampled values divided by the ESS.


\bsp	
\label{lastpage}
\end{document}